\documentclass[10pt, journal]{IEEEtran}

\usepackage{times}
\usepackage{epsfig}
\usepackage{graphicx}
\usepackage{amsmath}
\usepackage{amssymb}
\usepackage{algorithm}
\usepackage{algpseudocode}
\usepackage{epsfig}
\usepackage{subfigure}
\usepackage{lipsum}
\usepackage{booktabs}
\usepackage{multirow}
\usepackage{amsthm}
\usepackage{url}
\usepackage{breqn}
\usepackage{xcolor}
\usepackage{threeparttable}

\newcommand{\TheTitle}{Distributed Iterative CT Reconstruction using Multi-Agent Consensus Equilibrium
}

\title{\TheTitle}

\author{Venkatesh~Sridhar,~\IEEEmembership{Student Member,~IEEE,}
		Xiao ~Wang,~\IEEEmembership{ Member,~IEEE, }
		Gregery~T.~Buzzard,~\IEEEmembership{Member,~IEEE,}	
        Charles~A.~Bouman,~\IEEEmembership{Fellow,~IEEE}
\thanks{  
Venkatesh Sridhar and Charles A. Bouman are with the 
School of Electrical and Computer Engineering, 
Purdue University, 465 Northwestern Ave., West Lafayette,
IN 47907-2035, USA. 
Gregery T. Buzzard is with the Department of Mathematics, Purdue University, West Lafayette, IN, USA.
Xiao Wang is with the Boston Children's Hopsital and Harvard Medical School, Boston, MA, USA. 
Sridhar and Bouman were partially supported by the US Dept. of Homeland Security, S\&T Directorate, under Grant Award 2013-ST-061-ED0001.
Buzzard was partially supported by the NSF under grant award  CCF-1763896.
Email: 
\{vsridha,buzzard,bouman\}@purdue.edu, \{Xiao.Wang2\}@childrens.harvard.edu.}
}
    
\usepackage{amsopn}

\newtheorem{theorem}{Theorem}[section]

\newtheorem{lemma}[theorem]{Lemma}

\newcommand{\RR}{{\mathbb R}}

\newcommand{\argmin}{\operatorname*{argmin}}
\newcommand{\<}{\langle}
\renewcommand{\>}{\rangle}

\newcommand{\w}{{\mathbf w}}
\renewcommand{\u}{{\mathbf u}}
\newcommand{\x}{{\mathbf x}}
\newcommand{\y}{{\mathbf y}}
\newcommand{\z}{{\mathbf z}}
\renewcommand{\v}{{\mathbf v}}
\newcommand{\F}{{\mathbf F}}
\newcommand{\G}{{\mathbf G}}
\newcommand{\T}{{\mathbf T}}
\newcommand{\M}{{\mathbf M}}

\newcommand{\I}{{\mathbf I}}

\newcommand{\X}{{\mathbf X}}
\newcommand{\0}{{\bf 0}}

\renewcommand{\t}{{\mathbf t}}
\newcommand{\vs}{{\mathbf v}}
\newcommand{\cs}{{\mathbf c}}
\newcommand{\Ls}{{\mathbf L}}
\newcommand{\Lt}{\tilde{\Ls}}
\newcommand{\Hs}{{\mathbf H}}
\newcommand{\es}{\varepsilon_s}
\begin{document}



\maketitle
\thispagestyle{empty}

\begin{abstract}
Model-Based Image Reconstruction (MBIR) methods significantly enhance the quality of computed tomographic (CT) reconstructions relative to analytical techniques, but are limited by high computational cost.
In this paper, we propose a multi-agent consensus equilibrium (MACE) algorithm for distributing both the computation and memory of MBIR reconstruction across a large number of parallel nodes. 
In MACE, each node stores only a sparse subset of views and a small portion of the system matrix,
and each parallel node performs a local sparse-view reconstruction, which based on repeated feedback from other nodes, converges to the global optimum.
Our distributed approach can also incorporate advanced denoisers as priors to enhance reconstruction quality.
In this case, we obtain a parallel solution to the serial framework of Plug-n-play (PnP) priors, which we call MACE-PnP.
In order to make MACE practical, we introduce a partial update method that eliminates nested iterations and prove that it converges to the same global solution.  
Finally, we validate our approach on a distributed memory system with real CT data.
We also demonstrate an implementation of our approach on a massive supercomputer that can perform large-scale reconstruction in real-time.    

\end{abstract}


\section{INTRODUCTION}

Tomographic reconstruction algorithms can be roughly divided into two categories: analytical reconstruction methods \cite{FesslerBook14, Turbell01} and regularized iterative reconstruction methods \cite{Biester12} such as model-based iterative reconstruction (MBIR) \cite{Kisner12, DeMannBasu, BoumanICD, Sreehari}.
MBIR methods have the advantage that they can improve reconstructed image quality particularly when projection data are sparse and/or the X-ray dosage is low.
This is because MBIR integrates a model of both the sensor and object being imaged into the reconstruction process \cite{Kisner12, BoumanICD,Sreehari, Venkat}.
However, the high computational cost of MBIR often makes it less suitable for solving large reconstruction problems in real-time.

One approach to speeding MBIR is to precompute and store the system matrix \cite{Wang16, Sabne17, Matenine}. 
In fact, the system matrix can typically be precomputed in applications such as scientific imaging, non-destructive evaluation (NDE), and security scanning where the system geometry does not vary from scan to scan.
However, for large tomographic problems, the system matrix may become too large to store on a single compute node.
Therefore, there is a need for iterative reconstruction algorithms that can distribute the system matrix across many nodes in a large cluster.

More recently, advanced prior methods have been introduced into MBIR which can substantially improve reconstruction quality 
by incorporating machine learning approaches.
For example, Plug-n-play (PnP) priors \cite{Sreehari, Venkat}, consensus equilibrium (CE) \cite{Buzzard2018CE}, and 
RED \cite{Milanfar} allow convolutional neural networks (CNN) to be used in prior modeling.
Therefore, methods 
that distribute tomographic reconstruction across large compute clusters
should also be designed to support these emerging approaches. 

In order to make MBIR methods useful, it is critical to parallelize the algorithms for fast execution.
Broadly speaking, parallel algorithms for MBIR fall into two categories:
fine-grain parallelization methods that are most suitable for shared-memory (SM) implementation \cite{Wang16, Sabne17, Bouman2000, Fessler11},
and course-grain parallelization methods that are most suitable for distributed-memory (DM) implementation \cite{Wang17, Linyuan, Cui}.
So for example, SM methods are best suited for implementation on a single multi-core CPU processor or a GPU, 
while DM methods are better suited for computation across a large cluster of compute nodes.
Further, DM methods ensure that the overhead incurred due to inter-node communication and synchronization does not dominate the computation.

In particular, some DM parallel methods can handle large-scale tomographic problems by distributing the system-matrix across multiple nodes, while others do not.
For example, the DM algorithm of Wang et al. \cite{Wang17} parallelizes the reconstruction across multiple nodes, but it requires that each node have a complete local copy of the system matrix.
Alternatively, the DM algorithms of Linyuan et al. \cite{Linyuan} and Meng et al. \cite{Cui} could potentially be used to parallelize reconstruction across a cluster while distributing the system matrix across the nodes.
However, the method of Linyuan \cite{Linyuan} is restricted to the use of a total variation (TV) prior \cite{SauerBouman92, Chambolle} and in experiments has required 100s of iterations for convergence, which is not practical for large problems.
Alternatively, the the method of Meng \cite{Cui} is for use in unregularized PET reconstruction.

In this paper, we build on our previous work in \cite{Sridhar, WangSridhar} and present 
a Multi-agent Consensus Equilibrium (MACE) reconstruction algorithm that distributes both the computation and memory 
of iterative CT reconstruction across a large number of parallel nodes.
The MACE approach uses the consensus equilibrium \cite{Buzzard2018CE} framework to break the reconstruction problem into a set of subproblems which can be solved separately and then integrated together to achieve a high-quality reconstruction.
By distributing computation over a set of compute nodes, MACE enables the solution of reconstruction problems that would otherwise be too large to solve. 

\begin{figure*}[!ht]
    
     \begin{minipage}[l]{1.0\columnwidth}
         \centering
         \fbox{
         \includegraphics[width=0.97\textwidth]{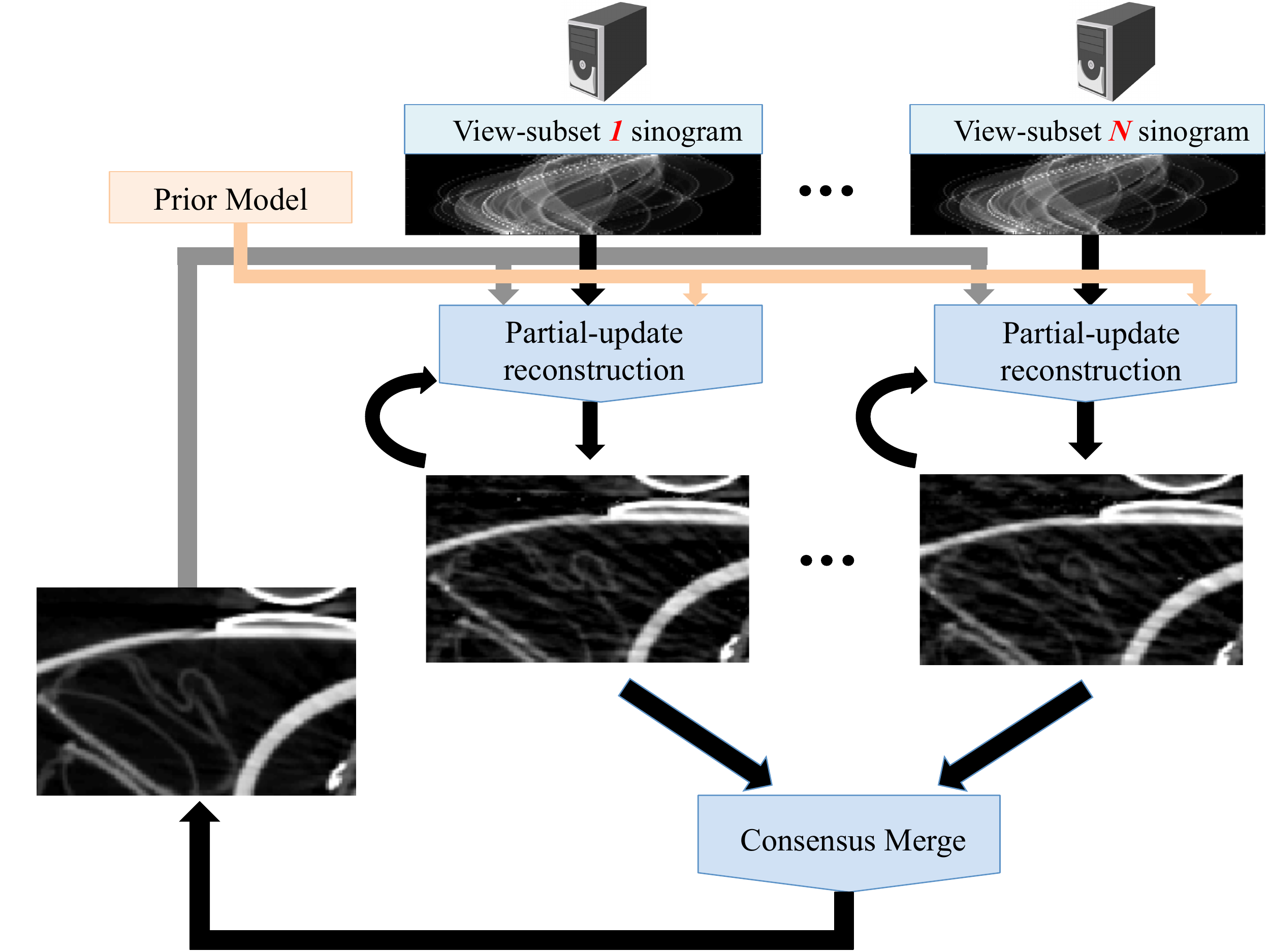}
         }
         \vspace{0.01in}
         \\ \small (a) MACE
     \end{minipage}
     \hspace{0.22in}
        \begin{minipage}[r]{1.0\columnwidth}
         \centering
         \fbox{
         \includegraphics[width=0.97\textwidth]{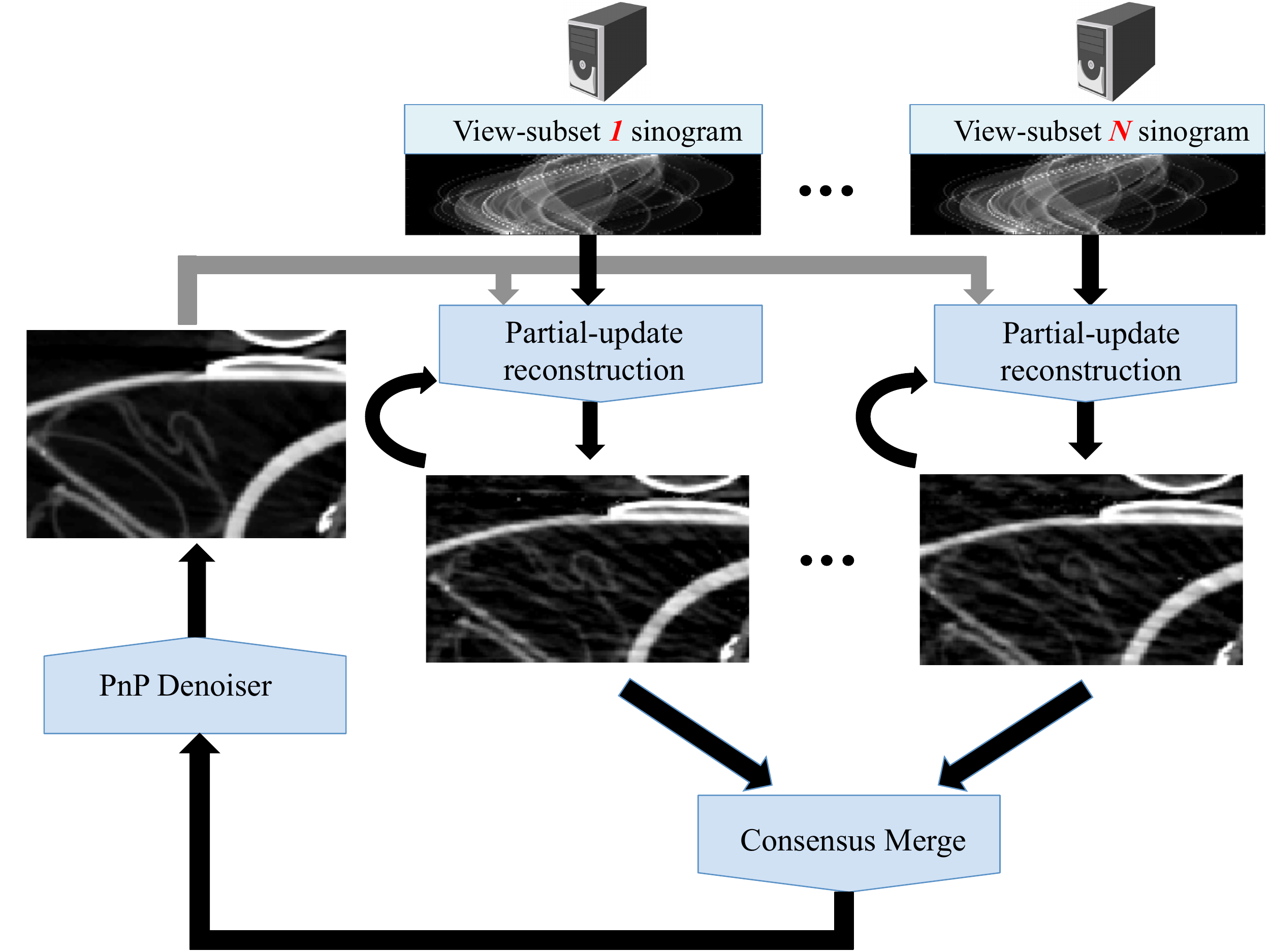}
         }
         \vspace{0.01in}
         \\ \small (b) MACE-PnP
        \end{minipage}

\caption{
Illustration of the MACE algorithm for distributing CT reconstruction across a parallel cluster of compute nodes. 
(a) An illustration of the MACE algorithm using a conventional prior model.
The MACE algorithm works by splitting the data into view subsets and reconstructing them in parallel.  
The individual reconstructions are them merged in an iterative loop that results in the true MAP reconstruction for the full data set.
(b) An illustration of the MACE-PnP algorithm which extends the MACE framework to the use of Plug-n-Play prior models to improve reconstruction quality.
In this case, a denoiser is run in the MACE loop in order to implement an advanced prior model.
The MACE and MACE-PnP algorithms distribute both computation and memory across parallel clusters of computers, thereby enabling the reconstruction of large tomographic data sets. 
}
\label{fig:MACEalgo}
\end{figure*}

Figure~\ref{fig:MACEalgo} illustrates our two approaches to this distributed CT reconstruction problem.
While both the approaches integrate multiple sparse-view reconstructions across a compute cluster into a high-quality reconstruction, they 
differ based on how the prior model is implemented.
Figure~\ref{fig:MACEalgo}(a) depicts our basic MACE approach that utilizes conventional edge-preserving regularization \cite{Kisner12, BoumanQGGMRF} as a prior model
and converges to the maximum-a-posteriori (MAP) estimate.
Figure~\ref{fig:MACEalgo}(b) shows our second approach called MACE-PnP which allows for distributed CT reconstruction using plug-and-play (PnP) priors \cite{Sreehari, Venkat}.
These PnP priors substantially improve reconstructed image quality by implementing the prior model using a denoising algorithm based on methods such as BM3D \cite{DabovBM3D07} or deep residual CNNs \cite{ZhangResnet}.
We prove that MACE-PnP provides a parallel algorithm for computing the standard serial PnP reconstruction of \cite{Sreehari}.

A direct implementation of MACE is not practical because it requires repeated application of proximal operators that are themselves iterative.
In order to overcome this problem, we introduce the concept of \emph{partial updates}, 
a general approach for replacing any proximal operator with a non-iterative update. 
We also prove the convergence of this method for our application.

Our experiments are divided into two parts and are based on real CT datasets from synchrotron imaging and security scanning.
In the first part, we use the MACE algorithm to parallelize 2D CT reconstructions across a distributed CPU cluster of 16 compute nodes.
We show that MACE both speeds up reconstruction while drastically reducing the memory footprint of the system matrix on each node. 
We incorporate regularization in the form of either conventional priors such as Q-GGMRF, or alternatively, advanced denoisers such as BM3D that improve reconstruction quality.
In the former case, we verify that our approach converges to the Bayesian estimate \cite{Kisner12, BoumanICD}, while in the latter case, we verify convergence to the PnP solution of \cite{Sreehari,Buzzard2018CE}.

In the second part of our experiments, we demonstrate an implementation of MACE on a large-scale supercomputer that can reconstruct a large 3D CT dataset.
For this problem, the MACE algorithm is used in conjunction with the super-voxel ICD (SV-ICD) algorithm \cite{Wang16} to distribute computation over 1200 compute nodes, consisting of a total of 81,600 cores.
Importantly, in this case the MACE algorithm not only speeds reconstruction,
it enables reconstruction for a problem that would otherwise be impossible since the full system matrix is too large to store on a single node.

\section{Distributed CT Reconstruction using Conventional Priors}
\label{sec:framework}
\subsection{CT Reconstruction using MAP Estimation}

We formulate the CT reconstruction problem using the maximum-a-posteriori (MAP) estimate given by \cite{BoumanSauerCT93} 
\begin{eqnarray}
x^* =  \argmin_{x \in \RR^n} f(x; \beta) \ , 
\label{eq:MAPest} 
\end{eqnarray}
where $f$ is the MAP cost function defined by
$$
f(x; \beta) = -\log p(y|x) - \log p (x) + const \ ,
$$
$y$ is the preprocessed projection data,
$x \in \RR^n$ is the unknown image of attenuation coefficients to be reconstructed,
and $const$ represents any possible additive constant.

For our specific problem, we choose the forward model $p(y|x)$ and the prior model $p_\beta (x)$ so that 
\begin{eqnarray}
f(x;\beta ) = \sum_{k=1}^{N_{\theta}}\frac{1}{2} \|y_k-A_k x\|_{\Lambda_k}^2 + \beta h(x) \ , 
\label{eq:global_obj}
\end{eqnarray}
where $y_k \in \RR^{N_D}$ denotes the $k^{th}$ view of data,
$N_{\theta}$ denotes the number of views,
$A_k \in \RR^{N_D \times n}$ is the system matrix that represents the forward projection operator for the $k^{th}$ view,
and $\Lambda_k \in \RR^{N_D \times N_D}$ is a diagonal weight matrix corresponding to the inverse noise variance of each measurement.
Also, the last term
$$
\beta h(x) = - \log p (x) + const \ ,
$$
represents the prior model we use where $\beta$ can be used to control the relative amount of regularization.
In this section, we will generally assume that $h(\cdot)$ is convex, so then $f$ will also be convex.
A typical choice of $h$ which we will use in the experimental section is the Q-Generalized Gaussian Markov Random Field (Q-GGMRF) prior \cite{BoumanQGGMRF} that preserves both low contrast characteristics as well as edges.


In order to parallelize our problem, we will break up the MAP cost function into a sum of auxiliary functions, 
with the goal of minimizing these individual cost functions separately.
So we will represent the MAP cost function as
$$
 f (x; \beta) = \sum_{i=1}^N  f_i(x; \beta) \ ,
$$
where 
\begin{align}
    f_i(x; \beta) =  \sum_{k \in J_i} \frac{1}{2} \|y_k-A_k x\|_{\Lambda_k}^2 + \frac{\beta}{N} h(x) \ ,
\label{eq:local_obj}
\end{align}
and the view subsets $J_1, \cdots , J_N$ partition the set of all views into $N$ subsets.\footnote{By partition, we mean that 
$\cup_{i=1}^N J_i = \{1,2,\cdots,N_{\theta}\}$ and $\cap_{i=1}^N J_i = \emptyset$.}
In this paper, we will generally choose the subsets $J_i$ to index interleaved view subsets, but the theory we develop works for any partitioning of the views.
In the case of interleaved view subsets, the view subsets are defined by  
$$
J_i = \left\{ \ m : m \text{ mod } N = i, \ m \in \{1,\cdots,N_{\theta}\} \ \right\} \ .
$$

\subsection{MACE framework}
\label{sec:MACE}

In this section, we introduce a framework, which we refer to as multi-agent consensus equilibrium (MACE) \cite{Buzzard2018CE},
for solving our reconstruction problem through the individual minimization of the terms $f_i (x)$ defined in (\ref{eq:local_obj}).\footnote{In this section, we suppress the dependence on $\beta$ for notational simplicity.}
Importantly, minimization of each function, $f_i (\cdot)$, has exactly the form of the MAP CT reconstruction problem but with the sparse set of views indexed by $J_i$. 
Therefore, MACE integrates the results of the individual sparse reconstruction operators, or agents, to produce a consistent solution to the full problem.

To do this, we first define the agent, or in this case the proximal map, for the $i^{th}$ auxiliary function as
\begin{align}
F_i(x ) = \argmin_{z \in \RR^n} \left\{f_i(z) + \frac{\|z-x\|^2}{2 \sigma^2} \right\} \ ,
\label{eq:prox_F}
\end{align}
where $\sigma$ is a user selectable parameter that will ultimately effect convergence speed of the algorithm.
Intuitively, the function $F_i(x )$ takes an input image $x$, and returns an image that reduces its associated cost function $f_i$ and is close to $x$.

Our goal will then be to solve the following set of MACE equations 
\begin{align}
\label{eq:CEsys_F}
F_i( x^*+u_i^*) &=x^* \mbox{\ for\ } i=1,...,N \ , \\
\label{eq:CEsys_u} 
\sum_{i=1}^N u_i^* &=0 \ , 
\end{align}
where $x^* \in \RR^n$ has the interpretation of being the consensus solution,
and each $u_i^*\in \RR^n$ represents the force applied by each agent that balances to zero.

Importantly, the solution $x^*$ to the MACE equations is also the solution to the MAP reconstruction problem of equation~(\ref{eq:MAPest}) (see Theorem 1 of~\cite{Buzzard2018CE}).
In order to see this, notice that since $f_i$ is convex we have that 
$$
\partial f_i(x^*) +  \frac{x^* - (x^*+u_i^*)}{\sigma^2} \ni 0, \ i=1,\cdots,N,
$$
where $\partial f_i$ is the sub-gradient of $f_i$.
So by summing over $i$ and applying (\ref{eq:CEsys_u}) we have that
$$
\partial f(x^*) \ni 0 \ .
$$
Which shows that $x^*$ is a global minimum to the convex MAP cost function.
We can prove the converse in a similar manner.

We can represent the MACE equilibrium conditions of (\ref{eq:CEsys_F}) and (\ref{eq:CEsys_u}) in a more compact notational form.
In order to do this, first define the stacked vector
\begin{align}
\v = \left[
\begin{array}{c}
v_1 \\
\vdots  \\
v_N
\end{array}
\right] \ ,
\end{align}
where each component $v_i \in \RR^n$ of the stack is an image.
Then we can define a corresponding operator $\F$ that stacks the set of agents as
 \begin{align}
\F(\v) = \left(
\begin{array}{c}
F_1(v_1) \\
\vdots  \\
F_N(v_N)
\end{array}
\right) \ ,
\label{eq:F_operators}
\end{align}
where each agent $F_i$ operates on the $i$-th component of the stacked vector.
Finally, we define a new operator $\G(\v)$ that computes the average of each component of the stack,
and then redistributes the result.
More specifically,
\begin{align}
\G(\v) = \left(
\begin{array}{c}
\bar{\v} \\
\vdots  \\
\bar{\v}
\end{array}
\right),
\label{eq:G_operators}
\end{align}
where $\bar{\v} = \frac{1}{N}\sum_{i=1}^N v_i$.

Using this notation, the MACE equations of~(\ref{eq:CEsys_F}) and~(\ref{eq:CEsys_u}) have the much more compact form of 
\begin{align}
    \F(\v^*) = \G(\v^*) \ ,
\label{eq:CEsys_compact}
\end{align}
with the solution to the MAP reconstruction is given by $x^*=\bar{\v}^*$ where $\bar{\v}^*$ is the average of the stacked components in $\v^*$.

The MACE equations of (\ref{eq:CEsys_compact}) can be solved in many ways,
but one convenient way to solve them is to convert these equations to a fixed-point problem,
and then use well known methods for efficiently solving the resulting fixed-point problem~\cite{Buzzard2018CE}.
It can be easily shown (see appendix A) that the solution to the MACE equations are exactly the fixed points of the operator $\T=(2\F-\I)(2\G-\I)$ given by
\begin{align}
\T \w^* =  \w^* \ ,
\label{eq:w*_fixed_pt}
\end{align}
where then $\v^* = (2\G-\I) \w^*$.

We can evaluate the fixed-point $\w^*$ using the \emph{Mann} iteration \cite{ParikhProximal}.
In this approach, we start with any initial guess and apply the following iteration,
\begin{equation}
\w^{(k+1)} = \rho \T \w^{(k)} + (1-\rho)\w^{(k)}, k\ge 0 \ ,
\label{eq:Mann}
\end{equation}
which can be shown to converge to $\w^*$ for $\rho \in (0,1)$.
The Mann iteration of (\ref{eq:Mann}) has guaranteed convergence to a fixed-point $\v^*$ if  $\T$ is non-expansive.
Both $F_i$, as well as its reflection, $2F_i-I$, are non-expansive, since each proximal map $F_i$ belongs to a special class of operators called \emph{resolvents} \cite{ParikhProximal}.
Also, we can easily show that $2\G-\I$ is non-expansive.
Consequently, $\T$ is also non-expansive and (\ref{eq:Mann}) is guaranteed to converge.

In practice, $\F$ is a parallel operator that evaluates $N$ agents that can be distributed over $N$ nodes in a cluster.
Alternatively, the $\G$ operator has the interpretation of a reduction operation across the cluster followed by a broadcast of the average across the cluster nodes.

\subsection{Partial Update MACE Framework}

A direct implementation of the MACE approach specified by (\ref{eq:Mann}) is not practical, since it requires a repeated application of proximal operators $F_i,  i=1,\cdots,N$, that are themselves iterative.
Consequently, this direct use of (\ref{eq:Mann}) involves many nested loops of intense iterative optimization, 
resulting in an impractically slow algorithm.

In order to overcome the above limitation, we propose a {\em partial-update MACE} algorithm that permits a faster implementation without affecting convergence.
In partial-update MACE, we replace each proximal operator with a fast non-iterative update 
in which we partially evaluate the proximal map, $F_i$, using only a single pass of iterative optimization.

Importantly, a partial computation of $F_i(\cdot \ ; \sigma)$ resulting from a single pass of iterative optimization will be dependent on the initial state.
So, we use the notation $\tilde{F}_i(\cdot \ ;\sigma,X_i)$ to represent the partial-update for $F_i(\cdot \ ; \sigma)$, where $X_i \in \RR^n$ specifies the initial state. 
Analogous to (\ref{eq:F_operators}), $\tilde{\F}(\v;\sigma,\X)$ then denotes the partial update for stacked operator $\F(\v;\sigma)$ from an initial state $\X \in \RR^{nN}$.  

Algorithm \ref{alg:puce_algo} provides the pseudo-code for the Partial-update MACE approach.
Note that this algorithm strongly resembles equation (\ref{eq:Mann}) that evaluates the fixed-point of map $(2\F-\I)(2\G-\I)$, except that $\F$ is replaced with its partial update as shown in line \ref{op:PUA_F} of the pseudo-code.
Note that this framework allows a single pass of any optimization technique to compute the partial update.
In this paper, we will use a single pass of the the Iterative Coordinate Descent (ICD) optimization method \cite{BoumanICD, Wang16}, that greedily updates each voxel, 
but single iterations of other optimization methods can also be used.

\makeatletter
\def\BState{\State\hskip-\ALG@thistlm}
\makeatother

\begin{algorithm}
\caption{Partial-update MACE with conventional priors} 
\label{alg:puce_algo}
\begin{algorithmic}[1]

\BState \emph{Initialize}:
\State $\w^{(0)} \gets \text{any value} \in \RR^{nN} $
\State $\X^{(0)} = \G(\w^{(0)})$
\State $k \gets 0$ 

\While{$\text{not converged}$}
\State $\v^{(k)} = \left( 2\G-\I \right) \w^{(k)}$
\State $\X^{(k+1)} = \tilde{\F}\left(\v^{(k)};\sigma,\X^{(k)} \right)$ \Comment{\scalebox{.75}{Approximate $\F(\v^{(k)})$}} \label{op:PUA_F}
\State $\w^{(k+1)} = 2\X^{(k+1)}-\v$ \Comment{ \scalebox{.75}{$\approx (2\F-\I)(2\G-\I)\w^{(k)}$}} \label{op:TV}
\State $\w^{(k+1)} \gets \rho \w^{(k+1)} + \left(1-\rho \right)\w^{(k)}$ \Comment{\scalebox{.75}{Mann update}} \label{op:Mann}
\State $k \gets k+1$
\EndWhile

\BState \emph{Solution}:
\State $x^* = \bar{\w}^{(k)}$ \Comment{\scalebox{.75}{Consenus solution}} \label{op:stop}
\end{algorithmic}
\end{algorithm}

In order to understand the convergence of the partial-update algorithm,
we can formulate the following update with an augmented state as
\begin{align}
\label{eq:PUCEupdate}
\begin{bmatrix}
\w^{(k+1)} \\
X^{(k+1)} \\
\end{bmatrix}
&=
\begin{bmatrix}
\rho & 0 \\
0 & 1 \\
\end{bmatrix}
\begin{bmatrix}
2\tilde{\F}\left( \vs^{(k)} ; X^{(k)} \right) - \vs^{(k)} \\
\tilde{\F}\left( \vs^{(k)}; X^{(k)} \right)  \\
\end{bmatrix} \\
&+
\begin{bmatrix}
1-\rho & 0 \\
0 & 0 \\
\end{bmatrix}
\begin{bmatrix}
\w^{(k)} \\
X^{(k)} \\
\end{bmatrix}, \nonumber
\end{align}
where $ \vs^{(k)}=(2\G-\I)\w^{(k)}$.  We specify $\tilde F$ more precisely in Algorithm~\ref{alg:pu_icd}.  
In Theorem \ref{thm:pu_fp}, we show that any fixed-point $(\w^*, \X^*)$ of this partial-update algorithm is a solution to the exact MACE method of (\ref{eq:w*_fixed_pt}).
Further, in Theorem \ref{thm:PU_convergence} we show that for the specific case where $f_i$ defined in (\ref{eq:local_obj}) is strictly quadratic, the partial-update algorithm has guaranteed convergence to a fixed-point.

\begin{theorem}\label{thm:pu_fp}
Let $f_i:\RR^n \rightarrow \RR, i=1,\cdots,N$ be a strictly convex and differentiable function.
Let $F_i$ denote the proximal map of $f_i$. 
Let $\tilde{F_i}(v;x)$ denote the partial update for $F_i(v)$ as specified by Algorithm \ref{alg:pu_icd}.
Then any fixed-point $(\w^*,\X^*)$ of the Partial-update MACE approach represented by (\ref{eq:PUCEupdate}) is a solution to the exact MACE approach specified by (\ref{eq:w*_fixed_pt}).
\end{theorem}

\begin{proof}
Proof is in Appendix B.
\end{proof}

\begin{theorem}\label{thm:PU_convergence}
Let $B$ be a positive definite $n \times n$ matrix, and let $f_i:\RR^n \rightarrow \RR^n, i=1,...,N$ each be given by
$$
f_i(x) =  \sum_{k \in J_i} \frac{1}{2} \|y_k-A_k x\|_{\Lambda_k}^2 + \frac{\beta}{N} x^TBx .
$$
Let $F_i$ denote the proximal map of $f_i$.
Let $\tilde{F}_i(v;x, \sigma)$, $v, x\in \RR^n$, denote the partial update for proximal operation $F_i(v ;\sigma)$ as shown in  Algorithm \ref{alg:pu_icd}.
Then equation (\ref{eq:PUCEupdate}) can be represented by a linear transform 
$$
\begin{bmatrix}
\w^{(k+1)} \\
X^{(k+1)} \\
\end{bmatrix}
=
\M_{\sigma^2}
\begin{bmatrix}
\w^{(k)} \\
X^{(k)} \\
\end{bmatrix}
+\cs(\y, \mathbf{A}, \rho),
$$
where $\M_{\sigma^2} \in \RR^{2Nn \times 2Nn}$ and $\cs \in \RR^{2Nn}$.
Also, for sufficiently small $\sigma > 0$, any eigenvalue of the matrix $\M_{\sigma^2}$ is in the range (0,1) and hence the iterates defined by (\ref{eq:PUCEupdate}) converge in the limit $k \rightarrow \infty$ to $( \w^*, \X^* ) $, the solution to the exact MACE approach specified by (\ref{eq:w*_fixed_pt}).

\end{theorem}

\begin{proof}
Proof is in Appendix B.
\end{proof}

Consequently, in the specific case of strictly convex and quadratic problems, 
the result of Theorem~\ref{thm:PU_convergence} shows that despite the partial-update approximation to the MACE approach, we converge to the exact consensus solution.
In practice, we use non-Gaussian MRF prior models for their edge-preserving capabilities, and so the global reconstruction problem is convex but not quadratic.
However, when the priors are differentiable, they are generically locally well-approximated by quadratics, and our experiments show that we still converge to the exact solution even in such cases.

\maketitle

\makeatletter
\def\BState{\State\hskip-\ALG@thistlm}
\makeatother

\begin{algorithm}
\caption{ICD-based Partial-update for proximal operation $F_i(v)$ using an initial state $x$} 
\label{alg:pu_icd}
\begin{algorithmic}[1]
\State Define $\es=[0,\cdots,1,\cdots,0]^t \in \RR^n$ \Comment{\scalebox{.75}{entry 1 at $s$-th position}} 
\State $z=x$ \Comment{\scalebox{.75}{Copy initial state}}  \label{op:inital}

\For{$s = 1$ to $n$}
\State $\alpha_s = \argmin_{\alpha} \left\{f_i(z+\alpha \es) + \frac{1}{2\sigma^2} \|z+\alpha \es -v\|^2 \right\} $  \label{op:alpha_update}
\State $z \gets z + \alpha_s \es $ \label{op:z_update}
\EndFor

\State $\tilde{F}_i(v;x) = z$ \Comment{\scalebox{.75}{Partial update}} \label{op:stop_icd}
\end{algorithmic}
\end{algorithm}

\section{MACE with Plug-and-Play Priors}

In this section, we generalize our approach to incorporate \emph{Plug-n-Play} (PnP) priors implemented with advanced denoisers \cite{Sreehari}.
Since we will be incorporating the prior as a denoiser, 
for this section we drop the prior terms of in equation~(\ref{eq:global_obj}) by setting $\beta=0$.
So let $f(x)=f(x;\beta=0)$ denote the CT log likelihood function of (\ref{eq:global_obj}) with $\beta=0$ and no prior term,
and let $F(x)$ denote its corresponding proximal map.
Then Buzzard et al. in \cite{Buzzard2018CE} show that the PnP framework of \cite{Sreehari} can be specified by the following equilibrium conditions 
\begin{align}
F(x^*-\alpha^* ; \ \sigma)&=x^* \label{eq:PnP_F} \\
H(x^*+\alpha^*) &=x^* \label{eq:PnP_H},
\end{align}
where $H: \RR^n \rightarrow \RR^n$ is the plug-n-play denoiser used in place of a prior model.
This framework supports a wide variety of denoisers including BM3D and residual CNNs that can be used to improve reconstruction quality as compared to conventional prior models \cite{Sreehari, Venkat}. 

Let $f_i (x) = f_i (x; \beta=0 )$ to be the log likelihood terms from~(\ref{eq:local_obj}) corresponding to the sparse view subsets,
and let $F_i (x)$ be their corresponding proximal maps.
Then in Appendix~C we show that the PnP result specified by (\ref{eq:PnP_F}) and (\ref{eq:PnP_H}) 
is equivalent to the following set of equilibrium conditions.
\begin{align}
F_i(x^*+u_i^* ; \ \sigma) &=x^*, \ i=1,\cdots,N, \label{eq:ParallelPnP_F} \\
H(x^*+\alpha^*) &=x^*       \label{eq:ParallelPnP_H}, \\
\sum_{i=1}^N u_i^* + \alpha^* &= 0 \label{eq:ParallelPnP_usum}.
\end{align}

Again, we can solve this set of balance equations by transforming into a fixed point problem.
One approach to solving equations~(\ref{eq:ParallelPnP_F}) \textendash \ (\ref{eq:ParallelPnP_usum})
is to add an additional agent, $F_{N+1}=H$, and use the approach of Section~\ref{sec:MACE} \cite{Majee}.
However, here we take a slightly different approach in which the denoising agent is applied in series, rather than in parallel.

In order to do this, we first specify a rescaled parallel operator $\F$ and a novel consensus operator $\G_H$, given by
\begin{align}
\F = \left(
\begin{array}{c}
F_1(v_1; \sqrt{N} \sigma)  \\
\vdots \\
F_N(v_N; \sqrt{N} \sigma)
\end{array}
\right)
\text{ and }
\G_H(\v) = \left(
\begin{array}{c}
H(\bar{\v}) \\
\vdots  \\
H(\bar{\v})
\end{array}
\right)
\label{eq:FG_PnP},
\end{align}
where $\bar{\v}=\sum_{i=1}^N v_i/N$.

In Theorem \ref{thm:FixedPtSolutionToPnP} below we show that we can solve the equilibrium conditions of (\ref{eq:ParallelPnP_F}) \textendash \ (\ref{eq:ParallelPnP_usum}) by finding the fixed-point of the map $\T_H=(2\F-\I)(2\G_H-\I)$.
In practice, we can implement the $G_H$ first computing an average across a distributed cluster, then applying our denoising algorithm,followed by broadcasting back a denoised, artifact-free version of the average.  

\begin{theorem}
Let $F_i, i=1,\cdots,N$, denote the proximal map of 
function $f_i$. 
Let maps $\F$ and $\G_H$ 
be defined by (\ref{eq:FG_PnP}).
Let $\hat{x}^*$ denote $N$ vertical copies of $x^*$.
Then, $(x^*, \u^*) \in \RR^n \times \RR^{nN}$ is a solution to the equilibrium conditions of (\ref{eq:ParallelPnP_F}) \textendash \ (\ref{eq:ParallelPnP_usum}) if and only if the point $\w^* = \hat x^* - N\u^*$ is a fixed-point of map $\T_H=(2\F-\I)(2\G_H-\I)$ and $\G_H(\w^*)=\hat{x}^*$.
\label{thm:FixedPtSolutionToPnP}
\end{theorem}
\begin{proof}
Proof is in Appendix C.
\end{proof}

When $\T_H$ is non-expansive, we can again compute the fixed-point $\w^*\in \RR^{nN}$ using the Mann iteration
\begin{equation}
\w^{(k+1)} = \rho (2\F-\I)(2\G_H-\I)\w^{(k)} + (1-\rho)\w^{(k)} \ .
\label{eq:Mann_PnP}
\end{equation} 
Then, we can compute the $x^*$ that solves the MACE conditions (\ref{eq:ParallelPnP_F}) \textendash \ (\ref{eq:ParallelPnP_usum}), or equivalently, the PnP conditions (\ref{eq:PnP_F}) \textendash \ (\ref{eq:PnP_H}), as $x^*=H\bar{\w}^*$.
Importantly, (\ref{eq:Mann_PnP}) provides a parallel approach to solving the PnP framework since the parallel operator $\F$ typically constitutes the bulk of the computation, as compared to consensus operator $\G_H$.

In the specific case when the denoiser $H$ is \emph{firmly non-expansive}, such as a proximal map, we show in Lemma \ref{lem:NonExpansivePnP} that $\T_H$ is non-expansive.
While there is no such guarantee for any general $H$, in practice, we have found that this Mann iteration converges.
This is consistent with previous experimental results that have empirically observed convergence of PnP \cite{Venkat,Buzzard2018CE} for a wide 
variety of denoisers including BM3D \cite{DabovBM3D07}, non-local means \cite{BuadesNLM}, or Deep residual CNNs \cite{ZhangResnet, Majee}.

\begin{lemma}
\label{lem:NonExpansivePnP}
If $\F$ and $\G_H$ are defined by (\ref{eq:FG_PnP}), and $H$ is firmly non-expansive,
then $\T_H=(2\F-\I)(2\G_H-\I)$ is non-expansive and the Mann iteration of (\ref{eq:Mann_PnP}) converges to the fixed point of $\T_H$. 
\label{lem:HG}
\end{lemma}

\begin{proof}
The proof is in Appendix C. 
\end{proof}

The Algorithm~\ref{alg:puce_algo_advpriors} below shows the partial update version of the Mann iteration from equation~(\ref{eq:Mann_PnP}).
This version of the algorithm is practical since it only requires a single update of each proximal map per iteration of the algorithm.

\makeatletter
\def\BState{\State\hskip-\ALG@thistlm}
\makeatother

\begin{algorithm}
\caption{Partial-update MACE with PnP priors} 
\label{alg:puce_algo_advpriors}
\begin{algorithmic}[1]

\BState \emph{Initialize}:
\State $\w^{(0)} \gets \text{any value} \in \RR^{nN} $
\State $\X^{(0)} = \G_H(\w^{(0)})$
\State $k \gets 0$ 

\While{$\text{not converged}$}
\State $\v^{(k)} = \left( 2\G_H-\I \right) \w^{(k)}$
\State $\X^{(k+1)} = \tilde{\F}\left(\v^{(k)};\sigma,\X^{(k)} \right)$ \Comment{\scalebox{.75}{Approximate $\F(\v^{(k)})$}} \label{op:PUA_F_adv}
\State $\w^{(k+1)} = 2\X^{(k+1)}-\v$ \Comment{ \scalebox{.75}{$\approx (2\F-\I)(2\G_H-\I)\w^{(k)}$}} \label{op:TV_adv}
\State $\w^{(k+1)} \gets \rho \w^{(k+1)} + \left(1-\rho \right)\w^{(k)}$ \Comment{\scalebox{.75}{Mann update}} \label{op:Mann_adv}
\State $k \gets k+1$
\EndWhile

\BState \emph{Solution}:
\State $x^* = H\bar{\w}^{(k)}$ \Comment{\scalebox{.75}{Consenus solution}} \label{op:stop_adv}
\end{algorithmic}
\end{algorithm}


\section{Experimental Results}


Our experiments are divided into two parts corresponding to 2D and 3D reconstruction experiments.
In each set of experiments, we analyze convergence and determine how the number of view-subsets affects the speedup and parallel-efficiency of the MACE algorithm.

Table~\ref{tab:Datasets}(a) lists parameters of the 2D data sets.
The first 2D data set was collected at the Lawerence Berkeley National Laboratory Advanced Light Source synchrotron and is one slice of a scan of a ceramic matrix composite material. 
The second 2D data set was collected on a GE Imatron multi-slice CT scanner and was reformated into parallel beam geometry.
For the 2D experiments, reconstructions are done with a image size of $512\times 512$, 
and the algorithm is implemented on a distributed compute cluster of 16 CPU nodes using the standard \emph{Message Parsing Interface} (MPI) protocol.
Source code for our distributed implementation can be found in \cite{MPIPackage}.

Table~\ref{tab:Datasets}(b) lists parameters of the 3-D parallel-beam CT dataset used for our supercomputing experiment.
Notice that for the 3D data set, the reconstructions are computed with an array size of $1280\times 1280$, effectively doubling the 
resolution of the reconstructed images.
This not only increases computation, but makes the system matrix 
larger, making reconstruction much more challenging.
For the 3D experiments, MACE is implemented on the massive NERSC supercomputer using 1200 multi-core CPU nodes belonging to the Intel Xeon Phi Knights Landing architecture, with 68 cores on each node.

\begin{table}[ht]
\centering

\caption{CT Dataset Description}
\label{tab:Datasets}
\subtable[2-D Datasets]
{
\begin{tabular}{|c|c|c|c|}
\hline
Dataset & \#Views & \#Channels & Image size \\
\hline     
\shortstack{\\  Low-Res.\  Ceramic\\   Composite (LBNL) } & 1024 & 2560 & 512 \!\!$\times$\!\! 512 \\
\hline
 &  &  &  \\[-1ex]
Baggage Scan (ALERT) & 720 & 1024 & 512 \!\!$\times$\!\! 512 \\
\hline
\end{tabular}
}
\medskip
\subtable[3-D Dataset]{
\begin{tabular}{|c|c|c|c|}
\hline
Dataset & \#Views & \#Channels & Volume size \\
\hline     
\shortstack{\\ High-Res.\ Ceramic  \\ Composite (LBNL) } & 1024 & 2560 & 1280 \!\!$\times$\!\! 1280 \!\!$\times$\!\! 1200 \\
\hline
\end{tabular}
}
\end{table}

\subsection{Methods}

For the 2D experiments, we compare our distributed MACE approach against a single-node method.
Both the single-node and MACE methods use the same ICD algorithm for reconstruction.
In the case of the single-node method, ICD is run on a single compute node that stores the complete set of views and the entire system matrix.
Alternatively, for the MACE approach computation and memory is distributed among $N$ compute nodes, with each node performing reconstructions using a subset of views.
The MACE approach uses partial updates each consisting of 1 pass of ICD optimization.

We specify the computation in units called \emph{Equits} \cite{Yu11}.
In concept, 1 equit is the equivalent computation of 1 full iteration of centralized ICD on a single node.
Formally, we define an equit as
$$
\text{\# Equits} = \frac{\mbox{(\# of voxel updates)}}{\mbox{(\# of voxels in ROI)*(\# of view subsets)}} \ .
$$
For the case of a single node, 1 equit is equivalent to 1 full iteration of the centralized ICD algorithm.
However, equits can take fractional values since non-homogeneous ICD algorithms can skip pixel updates or update pixels multiple times in a single iteration.
Also notice this definition accounts for the fact the the computation of an iteration is roughly proportional to the number of views being processed by the node.
Consequently, on a distributed implementation, 1 equit is equivalent to having each node perform 1 full ICD iteration using its subset of views.

Using this normalized measure of computation, the speedup due to parallelization is given by
$$
\text{Speedup(N)} = N \times \frac{\text{(\# of equits for centralized convergence)}}{\text {(\# of equits for MACE convergence)}} \ ,
$$
where again $N$ is the number of nodes used in the MACE computation.
From this we can see that the speedup is linear in $N$ when the number of equits required for convergence is constant.

In order to measure convergence of the iterative algorithms, we define the NRMSE metric as
$$
\text{NRMSE}(x,x^*) = \frac{\|x-x^*\|}{\|x^*\| }.
$$
where $x^*$ is the fully converged reconstruction.


All results using the 3D implemented on the NERSC supercomputer used the highly parallel 3-D Super-voxel ICD (SV-ICD) algorithm described in detail in \cite{Wang16}.
The SV-ICD algorithm employees a hierarchy of parallelization to maximize utilization of each cluster nodes.
More specifically, the $1200$ slices in the 3D data set are processed in groups of 8 slices, with each group of 8 slices being processed by a single node.
The 68 cores in each node then perform a parallelized version of ICD in which pixels are processed in blocks called super-voxels.
However, even with this very high level of parallelism, the SV-ICD algorithm has two major shortcomings.
First, it can only utilize 150 nodes in the super-computing cluster, but more importantly, the system matrix for case is too large to store on a single node,
so high resolution 3D reconstruction is impossible without the MACE algorithm.

\begin{figure}
\centering
\subfigure{\includegraphics[width =0.24\textwidth]{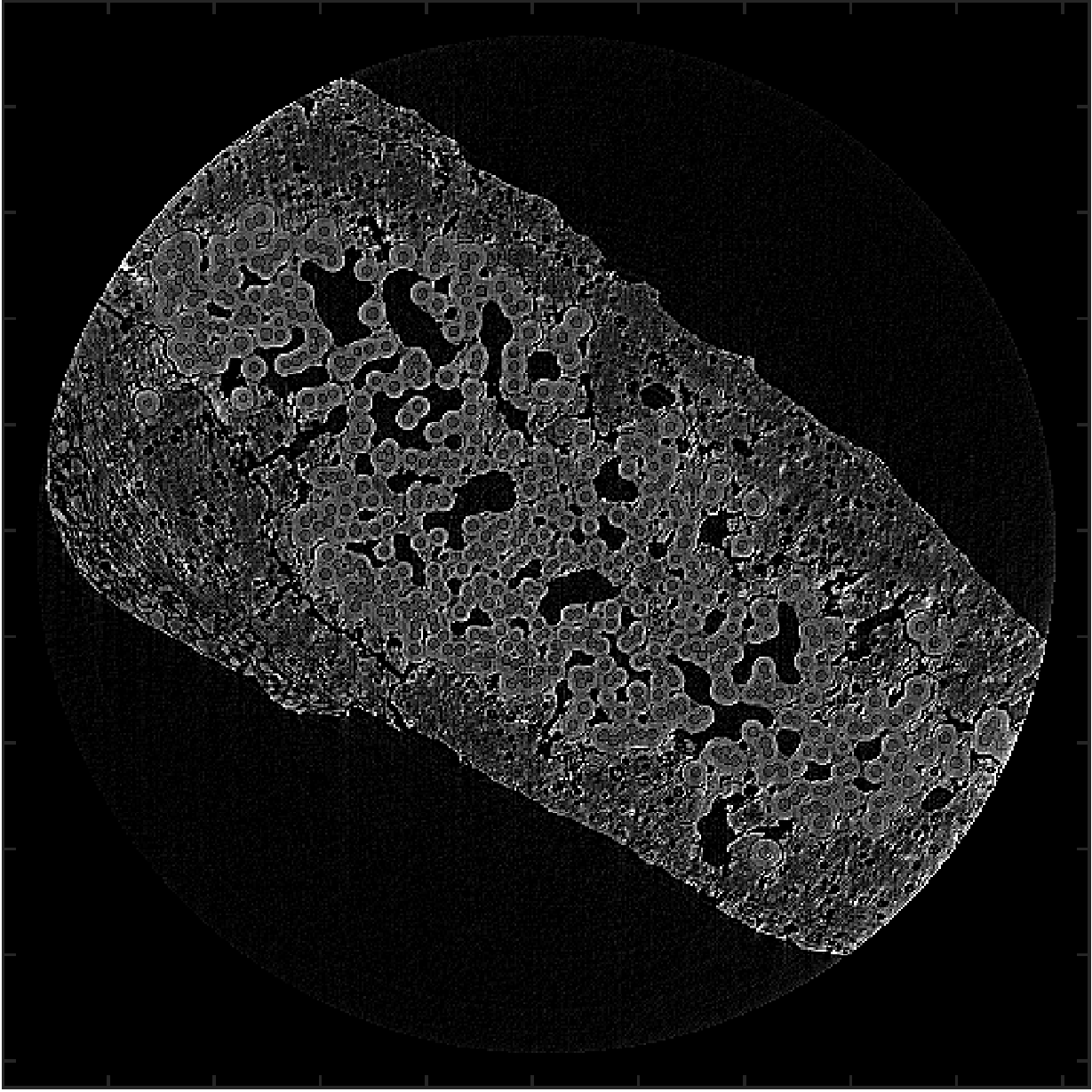}} 
\subfigure{\includegraphics[width=0.24\textwidth]{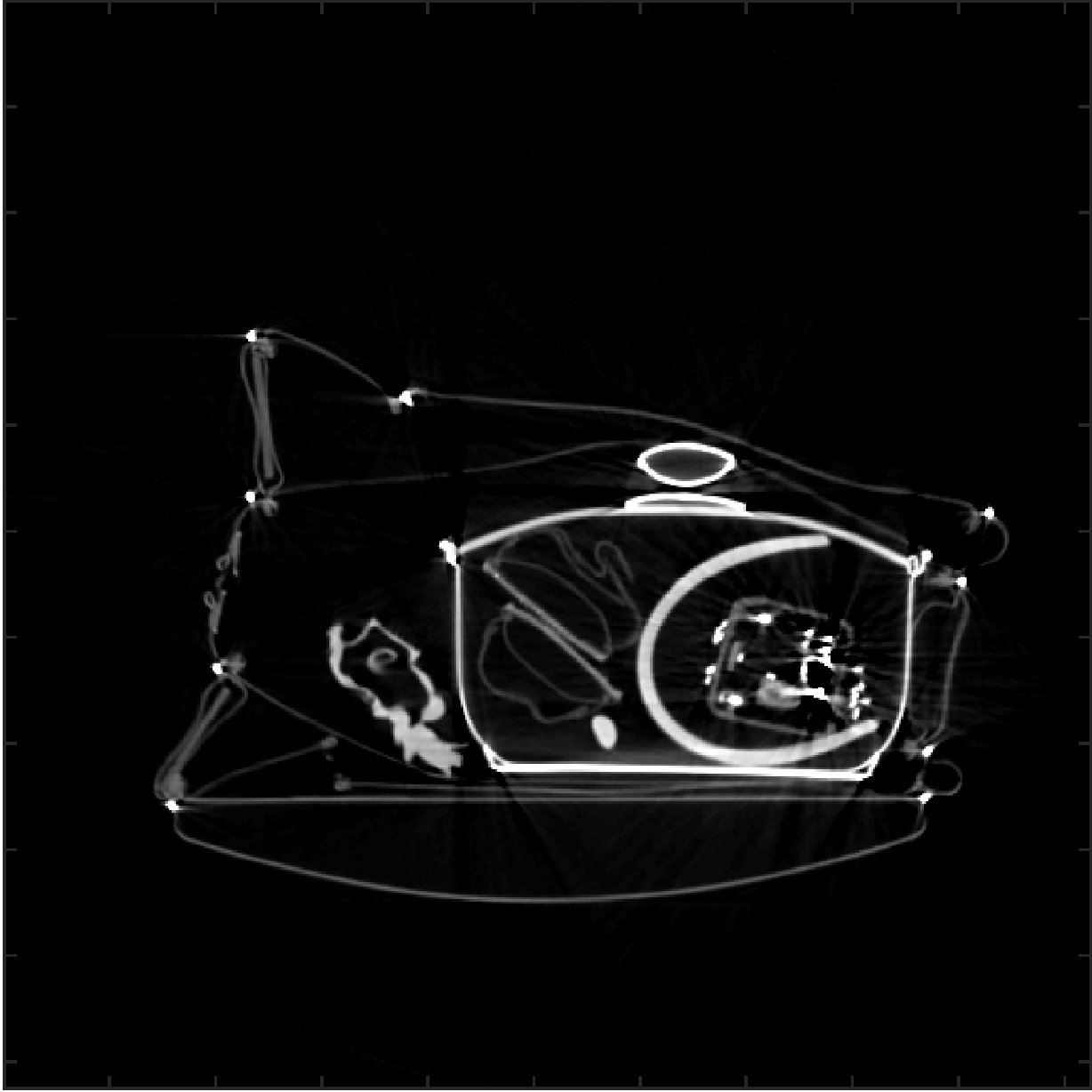}} 
\caption{
Single node reconstruction for (left) Low-Res. Ceramic Composite dataset (right) Baggage Scan dataset. 
}
\label{fig:centralized}
\end{figure}

\begin{figure}
\centering
\subfigure[Single node reconstruction ]{\includegraphics[width =0.24\textwidth]{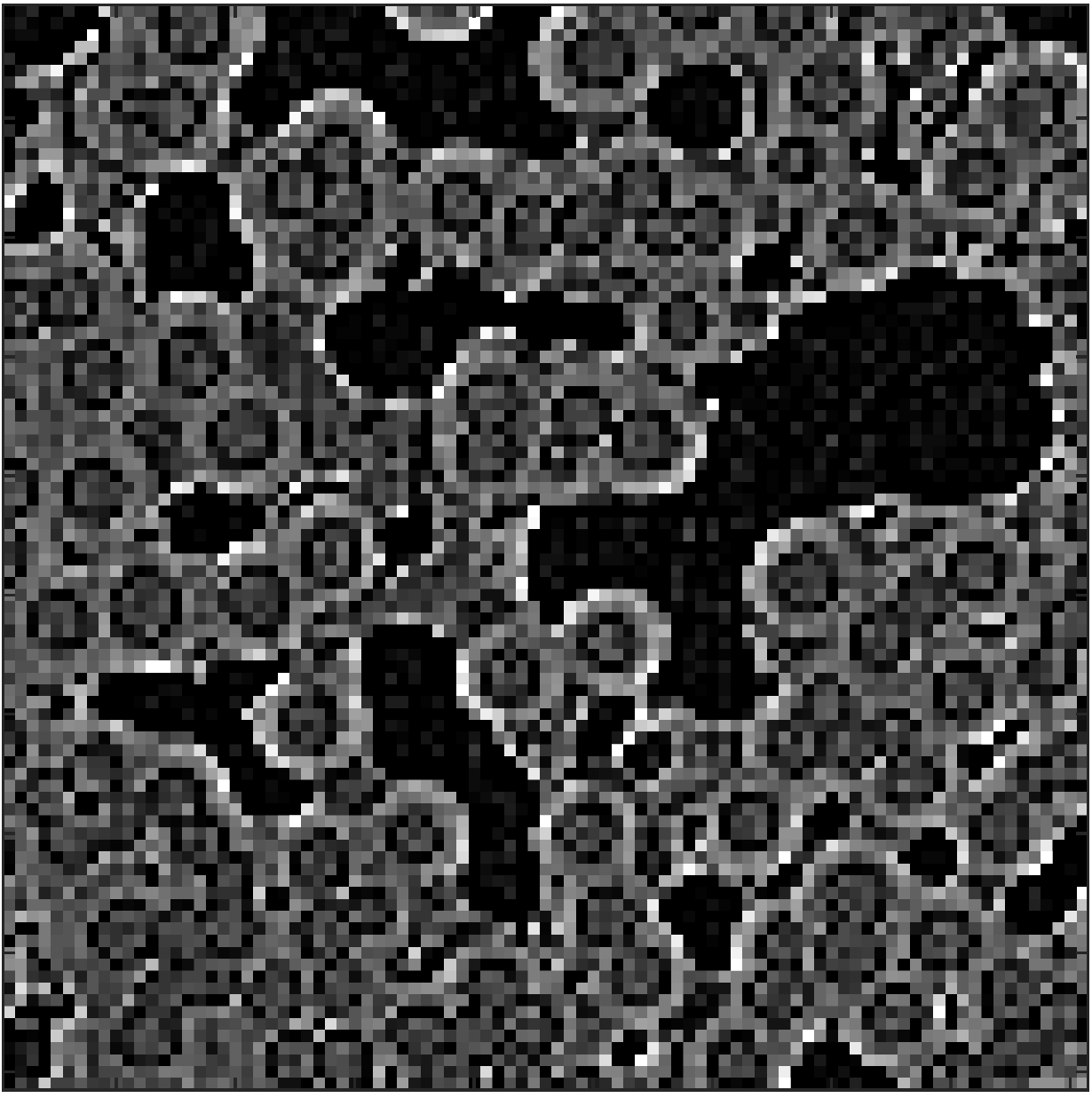}} 
\subfigure[MACE reconstruction, $N=16$]{\includegraphics[width =0.24\textwidth]{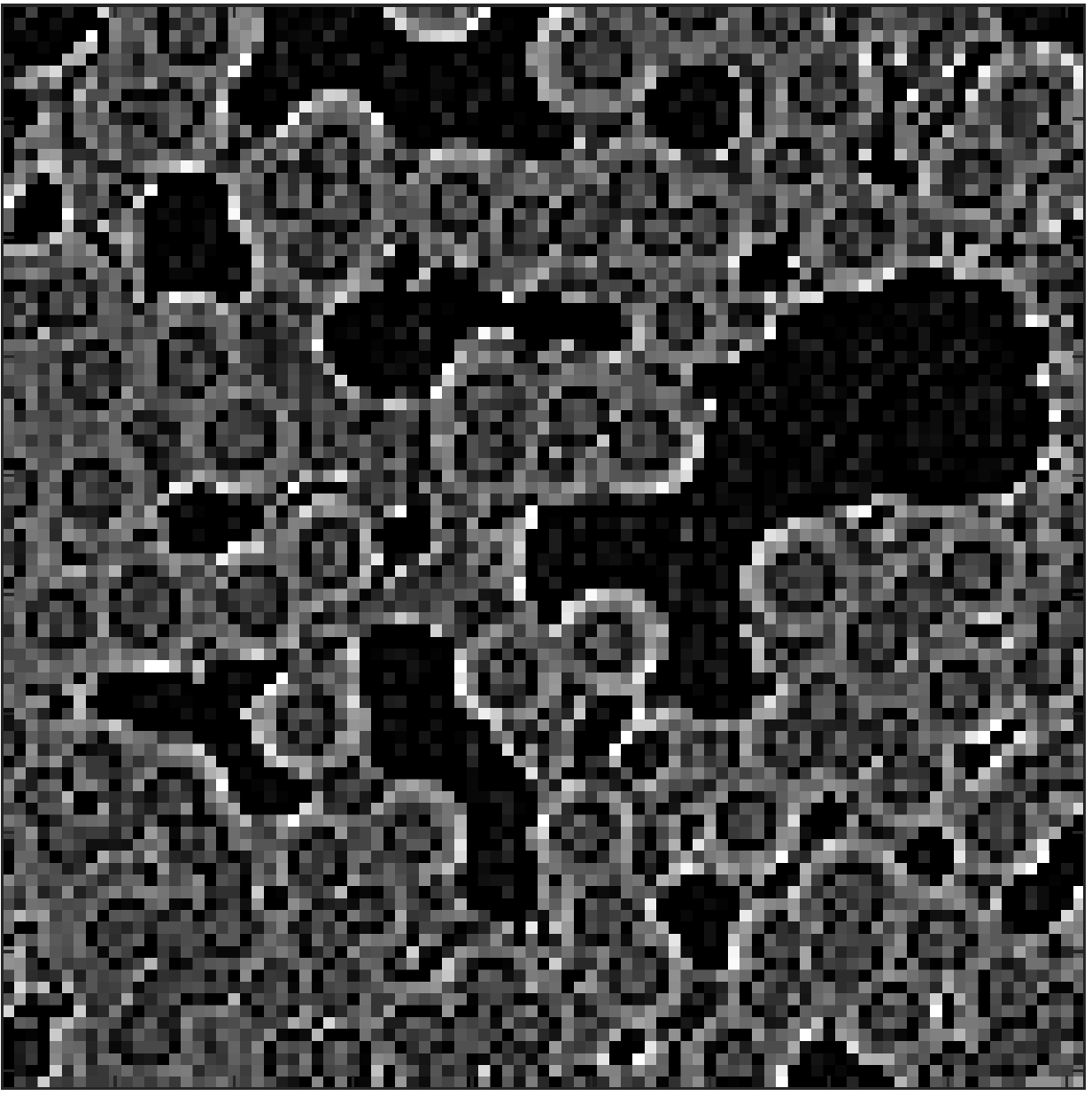}} 
\subfigure[Single node reconstruction]{\includegraphics[width =0.24\textwidth]{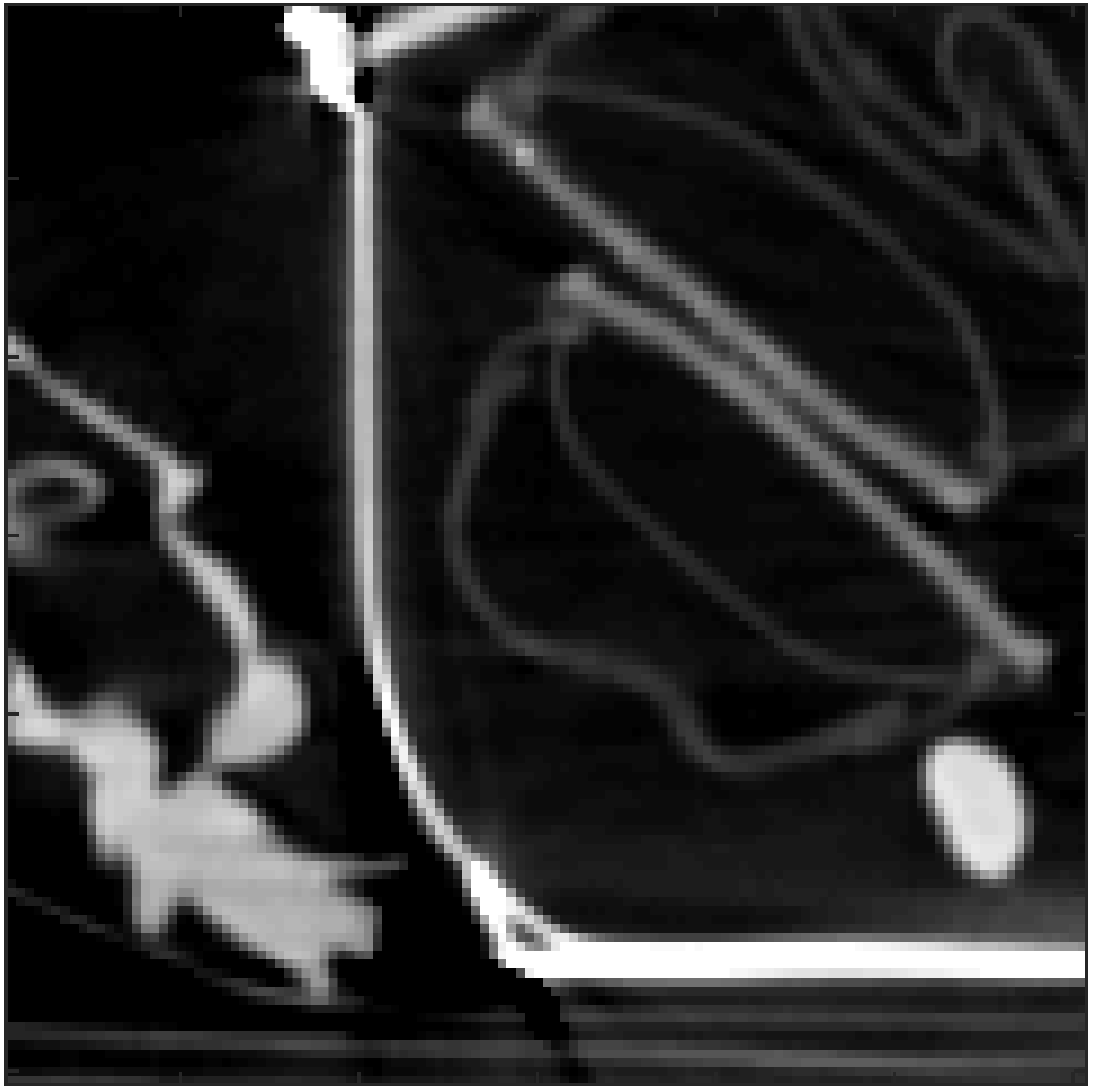}}
\subfigure[MACE reconstruction, $N=16$]{\includegraphics[width =0.24\textwidth]{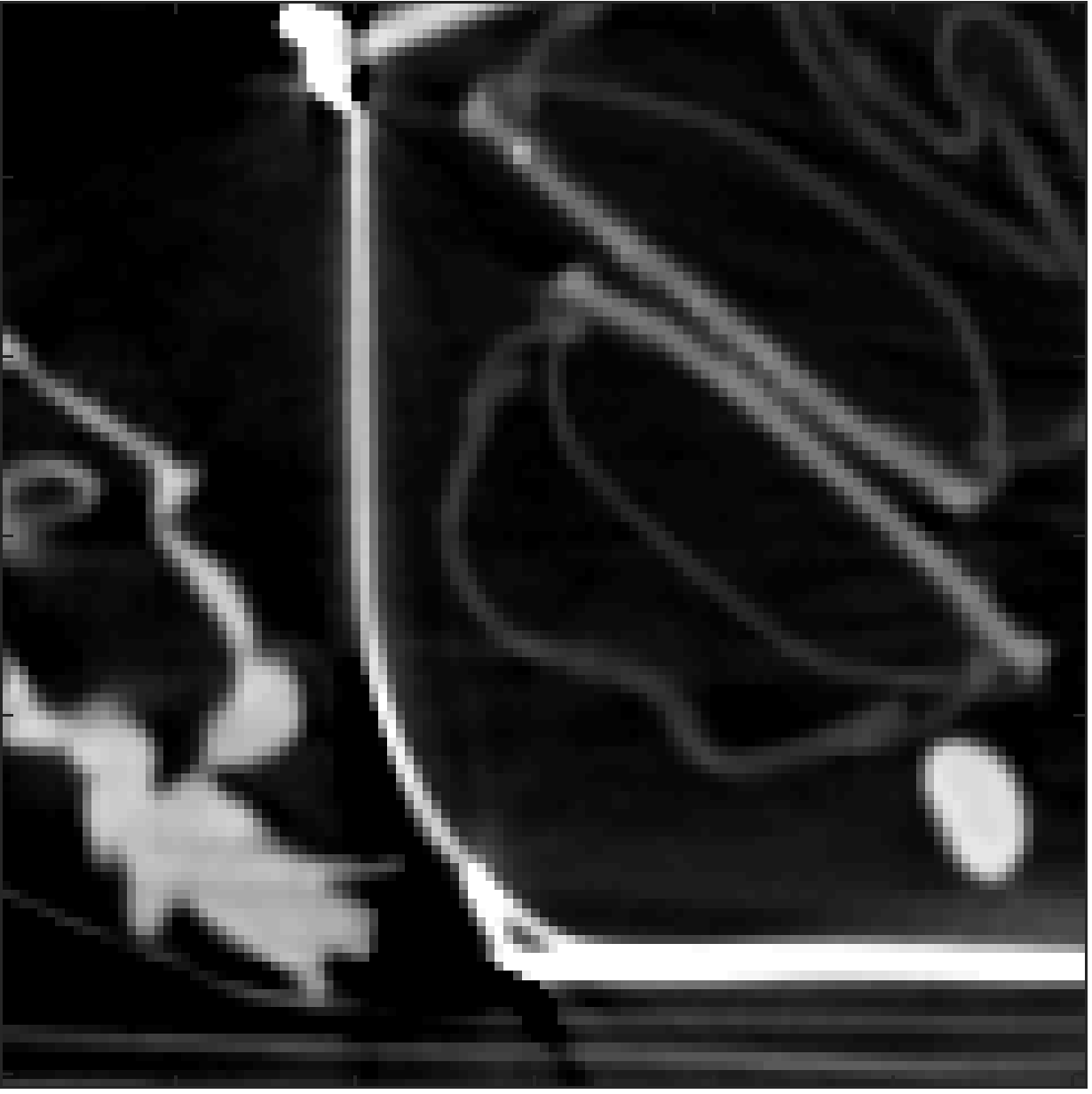}} 

\caption{
Comparison of reconstruction quality for MACE method using 16 parallel nodes each processing $(1/16)^{th}$ of the views, against centralized method.
(a) and (c) Centralized method. (b) and (d) MACE.
Notice that both methods have equivalent image quality. 
}
\label{fig:MACE_comparison}
\end{figure}

\begin{table}
\centering
\small
\caption{MACE convergence (equits) for different values of $\rho$, $N$=16.}
\begin{tabular}{p{2.7cm} p{0.65cm} p{0.65cm} p{0.65cm} p{0.65cm} p{0.65cm}}
\toprule
\text{Dataset} & \multicolumn{5}{c@{}}{$\rho$}\\
    \cmidrule(l){2-6}
    & 0.5 & 0.6 & 0.7 & 0.8 & 0.9\\
\midrule
Low-Res. Ceramic & 18.00 & 16.00 & 15.00 & \textbf{14.00} & 14.00\\
\hline
Baggage scan & 12.64 & 10.97 & 10.27 & \textbf{9.52} & 12.40\\
\bottomrule
\end{tabular}

\label{tab:Mann}
\end{table}

\begin{figure}
\centering
\subfigure[Convergence for different \#nodes, Low-Res. Ceramic dataset]{\includegraphics[width=0.42\textwidth]{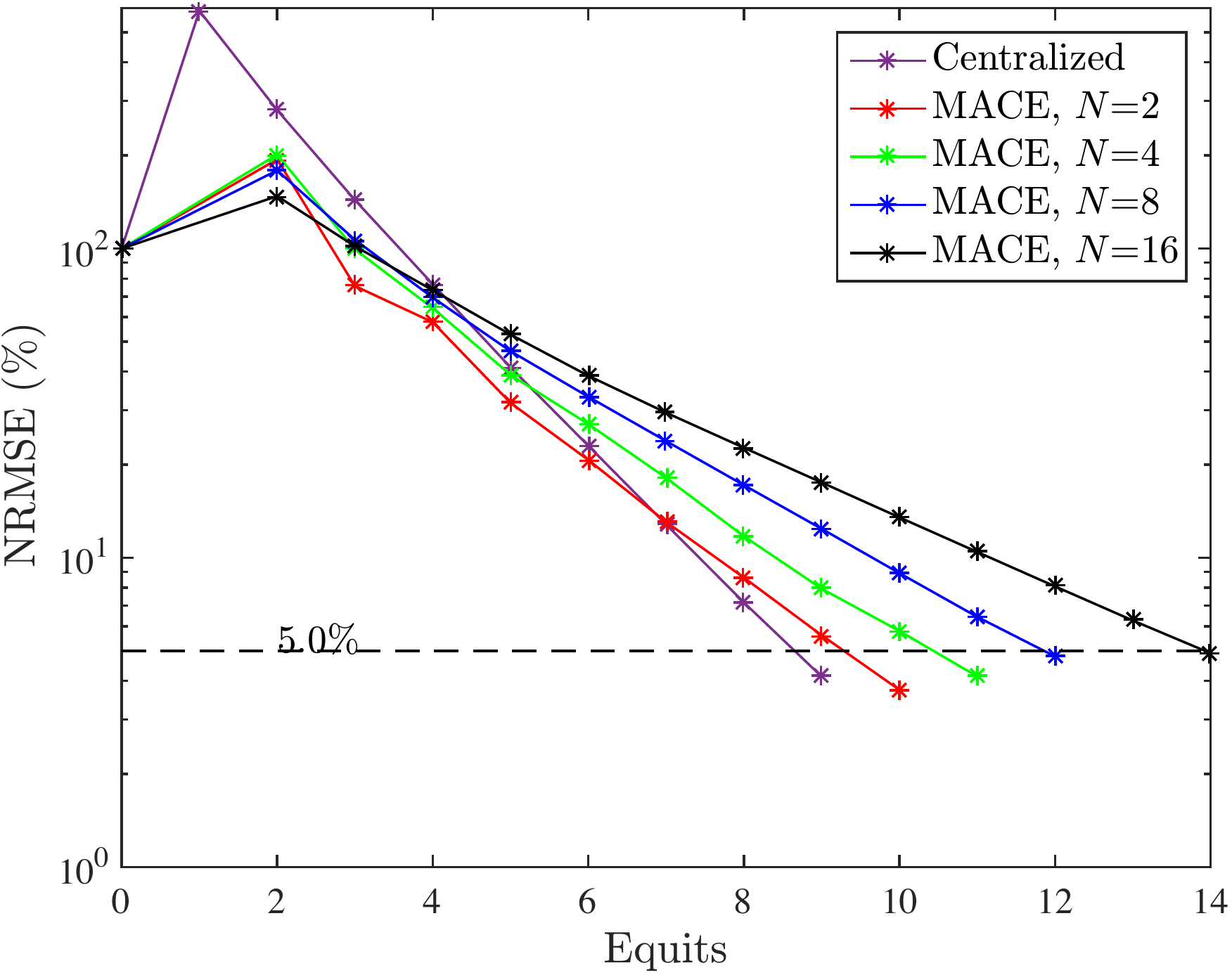}} 
\subfigure[Convergence for different \#nodes, Baggage Scan dataset]{\includegraphics[width=0.42\textwidth]{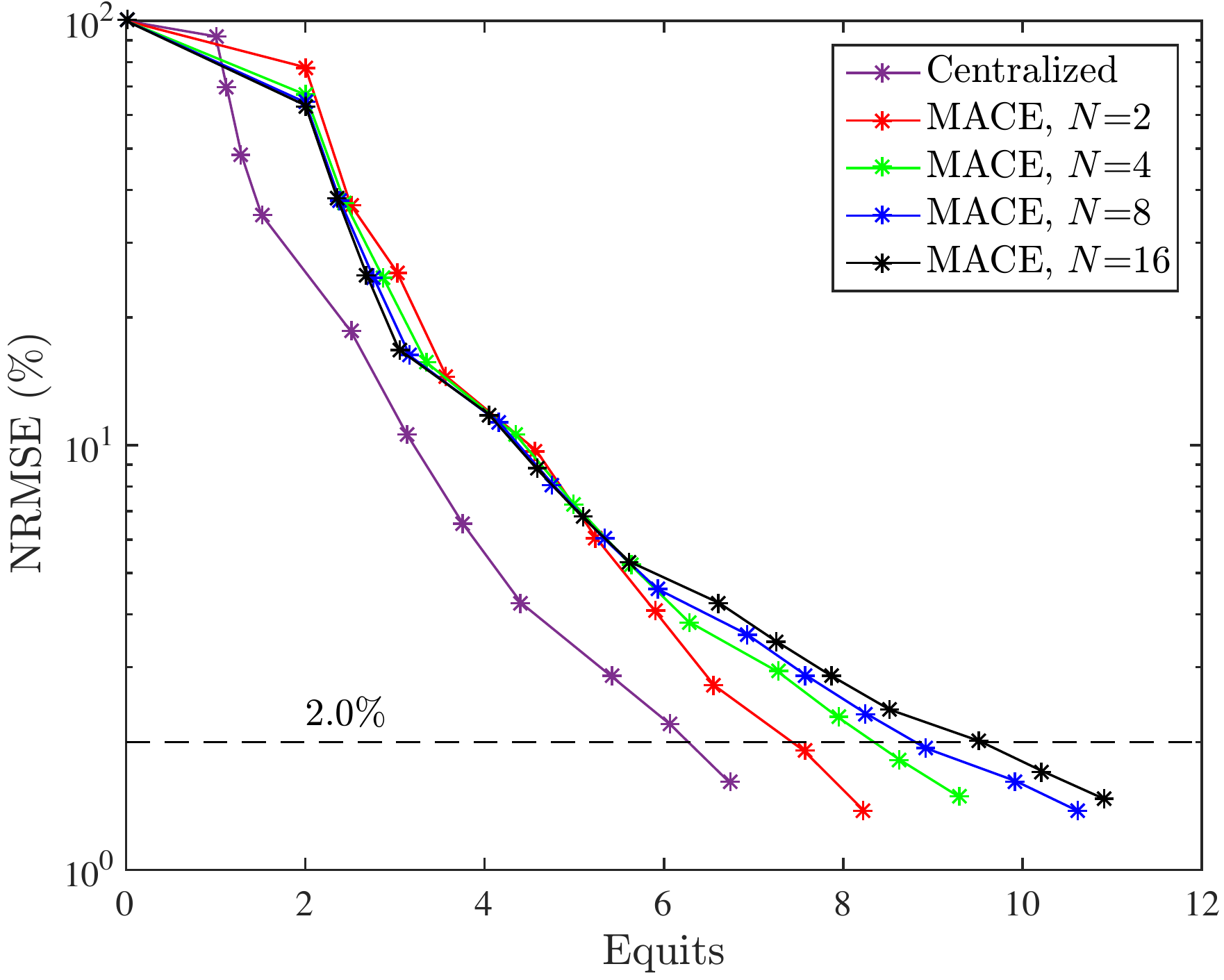}} 
\caption{
MACE convergence for different number of nodes, $N$, using $\rho=0.8$:
(a) Low-Res. Ceramic composite dataset
(b) Baggage Scan dataset.
Notice that number of equits tends to gradually increase with number of parallel processing nodes, $N$.
}
\label{fig:NumNodes_plots}
\end{figure}

\begin{figure}
\centering
\subfigure[16 node MACE reconstruction with PnP prior]{\includegraphics[width=0.40\textwidth]{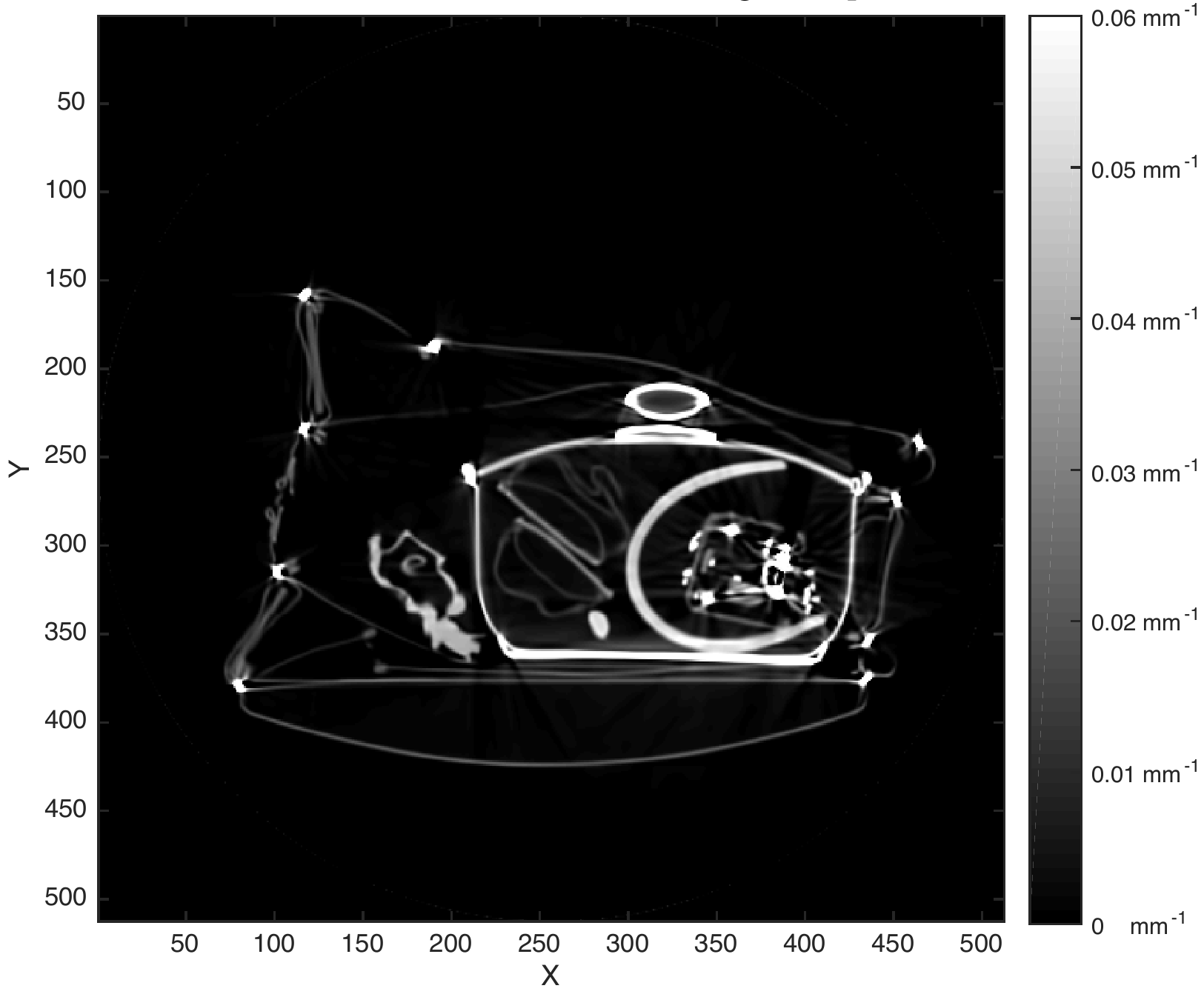}} 
\subfigure[Comparison of PnP prior (left) vs conventional prior (right)]{\includegraphics[width=0.42\textwidth]{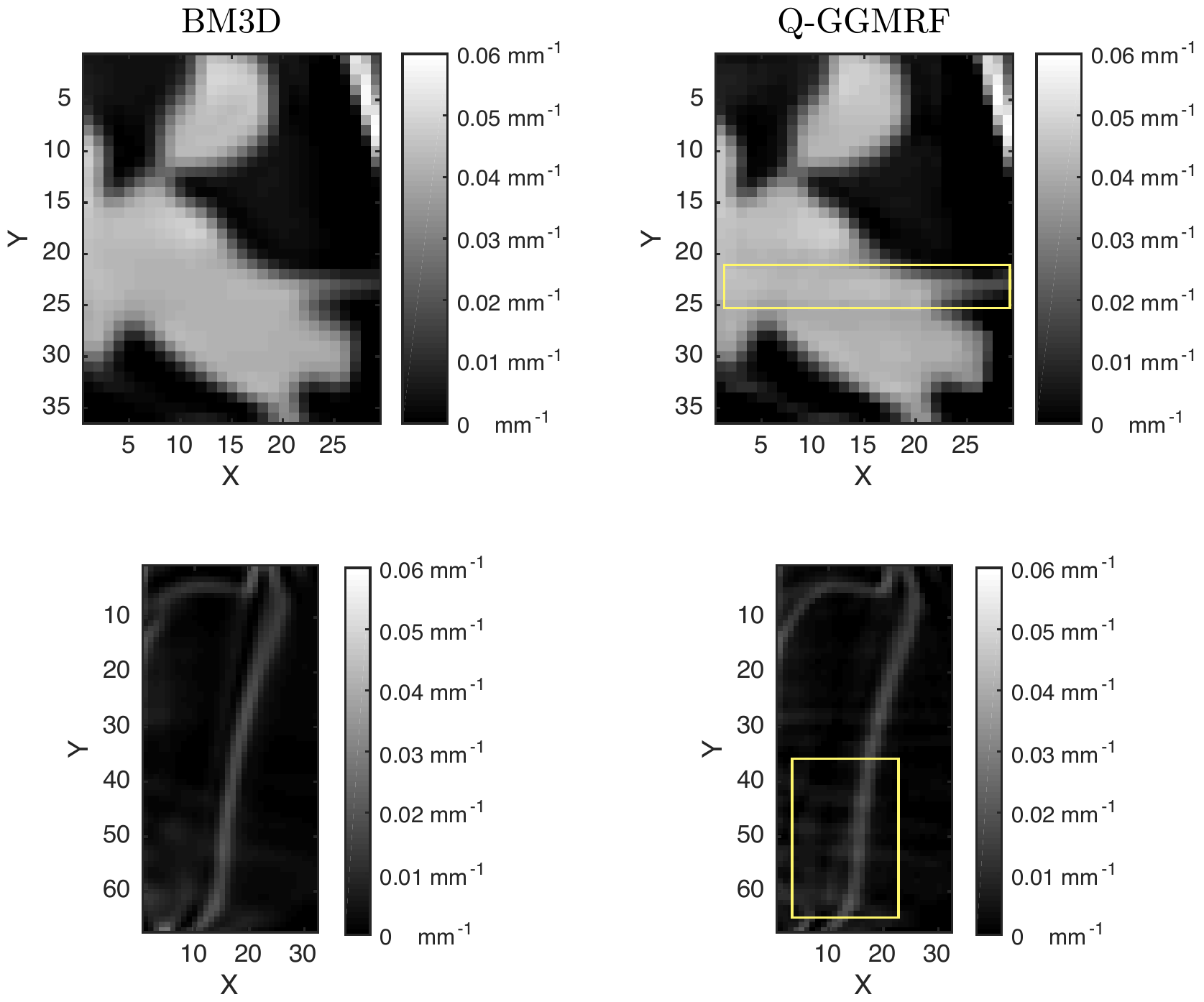}} 
\subfigure[Convergence for different \#nodes]{\includegraphics[width=0.42\textwidth]{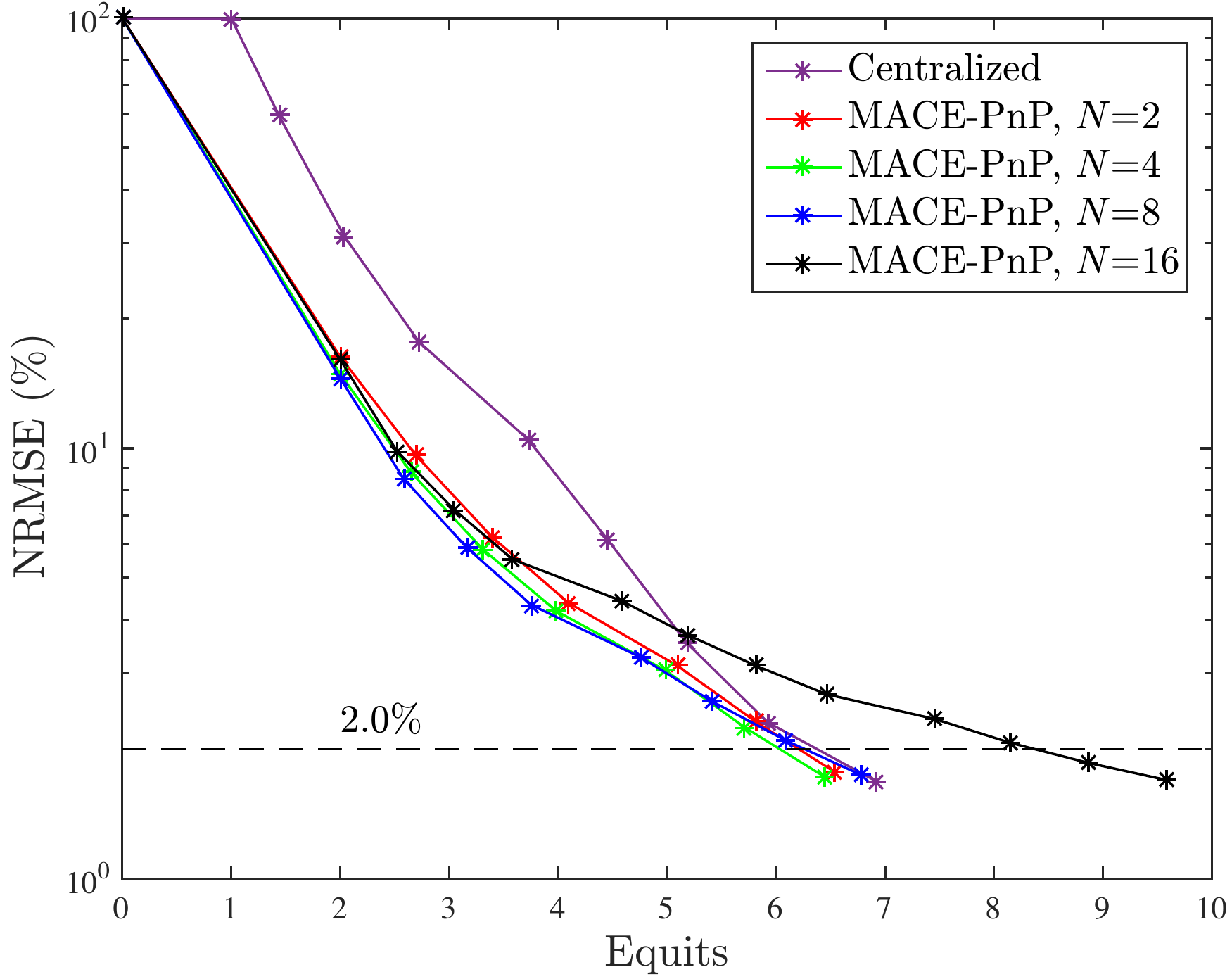}} 
\caption{
MACE-PnP reconstruction of \mbox{Baggage Scan} data set: 
(a) MACE PnP reconstruction using $N=16$ nodes;
(b) Zoomed-in regions of PnP versus conventional prior;
(c) MACE-PnP convergence as a function of number of nodes $N$.
Notice that PnP prior produces better image quality with reduced streaking and artifacts.
In addition, the number of equits required for convergence of the MACE-PnP does not tend to increase significantly with number of nodes $N$.
}
\label{fig:alert_pnp_img}
\end{figure}

\begin{figure}[ht]
    \centering
    \includegraphics[width=0.48\textwidth]{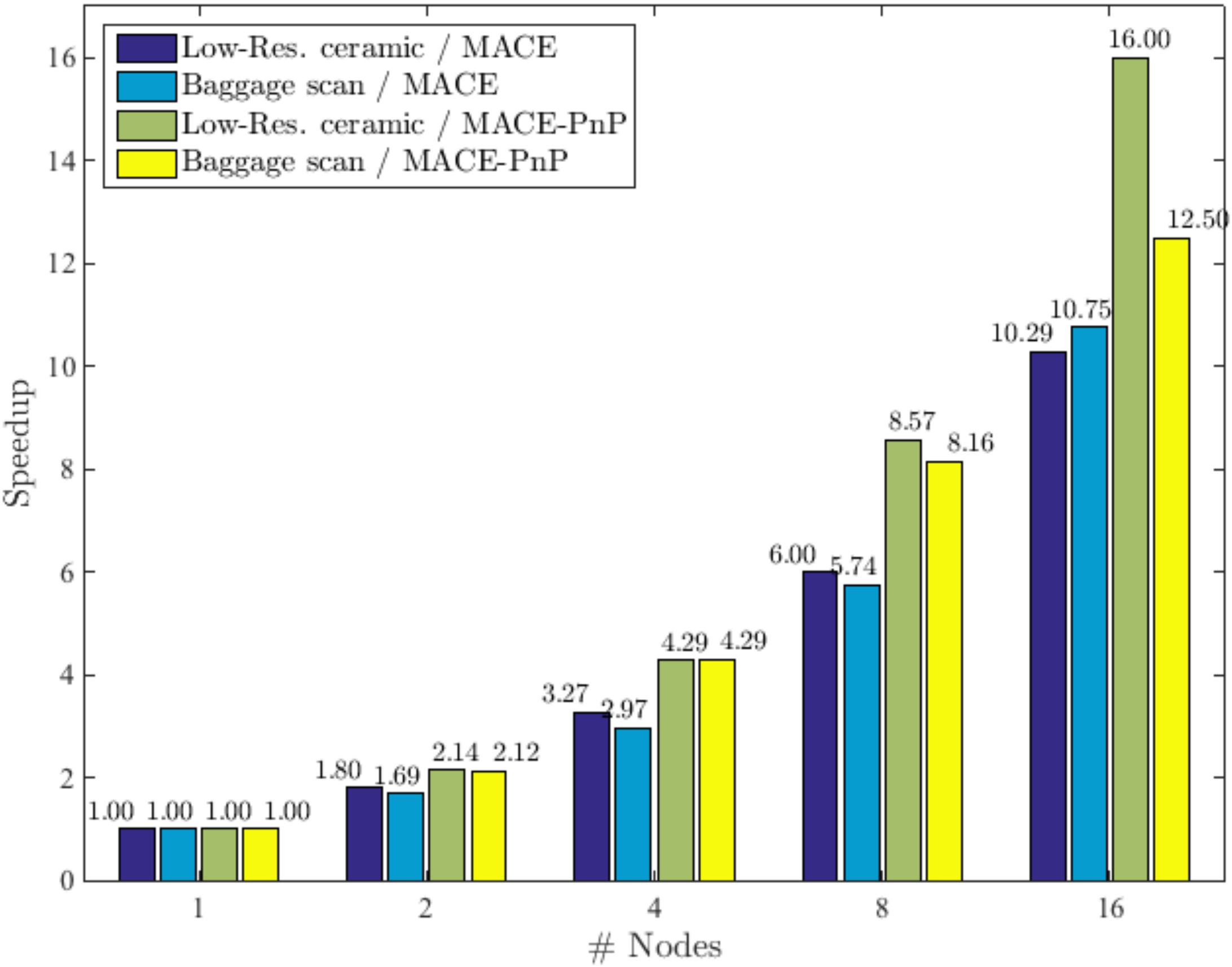}
    \caption{
    MACE speedup as a function of the number of nodes, for different datasets and prior models. 
    Importantly, note that in the case of PnP priors, we achieve a near linear speedup for both datasets.
    }
    \label{fig:Speedup_dsets}
\end{figure}


\subsection{MACE Reconstruction of 2-D CT Dataset}

In this section, we study the convergence and parallel efficiency of the 2D data sets of Table~\ref{tab:Datasets}(a).
Figure~\ref{fig:centralized} shows reconstructions of the data sets,
and Figure~\ref{fig:MACE_comparison} compares the quality of the centralized and MACE reconstructions for zoomed-in regions of the image.
The MACE reconstruction is computed using $N=16$ compute nodes or equivalently $N=16$ view-subsets.
Notice that the MACE reconstruction is visually indistinguishable from the centralized reconstruction.
However, for the MACE reconstruction, each node only stores less than $7 \%$ of the full sinogram, dramatically reducing storage requirements.

Table~\ref{tab:Mann} shows the convergence of MACE for varying values of the parameter $\rho$, with $N$=16.
Notice that for both data sets, $\rho = 0.8$ resulted in the fastest convergence.
In fact, in a wide range of experiments, we found that $\rho = 0.8$ was a good choice and typically resulted in the fastest convergence.

Figures~\ref{fig:NumNodes_plots} shows the convergence of MACE using a conventional prior for varying numbers of compute nodes, $N$. 
Notice that for this case of the conventional QGGMRF prior model, as $N$ increased, the number of equits required for convergence tended to increase for both data sets.

Figure~\ref{fig:alert_pnp_img} shows results using the MACE with the PnP prior (MACE-PnP) with the \mbox{Baggage Scan} data set.
Notice that the PnP prior results in improved image quality with less streaking and fewer artifacts.
Also notice that with the PnP prior, the number of equits required for convergence shown in Figure~\ref{fig:alert_pnp_img}(c) does not significantly increase with number of compute nodes, $N$.

Figure~\ref{fig:Speedup_dsets} summarizes this important result by plotting the parallel speed up as a function of number of nodes for both data sets using both priors.
Notice that the MACE-PnP algorithm results in approximately linear speedup for $N\leq8$ for both data sets, and near linear speedup up to $N=16$.
We conjecture that the advance prior tends to speed convergence due the stronger regularization constraint it provides.

Table~\ref{tab:Memory_MPI} shows the system matrix memory as a function of the number of nodes, $N$.
Note that MACE drastically reduces the memory usage per node by a factor of $N$.

\begin{table}
\centering
\small
\begin{threeparttable}
\caption{MACE memory usage for the system-matrix (gigabytes) as a function of number of nodes, $N$ \tnote{1}}
\label{tab:Memory_MPI}
\begin{tabular}{p{2.7cm} p{0.65cm} p{0.65cm} p{0.65cm} p{0.65cm} p{0.65cm}}
\toprule
\text{Dataset} & \multicolumn{5}{c@{}}{$N$}\\
    \cmidrule(l){2-6}
    & 1 & 2 & 4 & 8 & 16\\
\midrule
Low-Res. Ceramic & 14.83 &  7.42 &  3.71 &  1.86 & 0.93\\
\hline
Baggage scan & 5.04 &  2.52 &  1.26 &  0.63 & 0.32 \\
\bottomrule
\end{tabular}
\begin{tablenotes}
\item[1] \footnotesize{System-matrix represented in sparse matrix format and floating-point precision} 
\end{tablenotes}
\end{threeparttable}

\end{table}


\begin{figure}
\centering

\subfigure[Fully converged]{\includegraphics[width =0.22\textwidth]{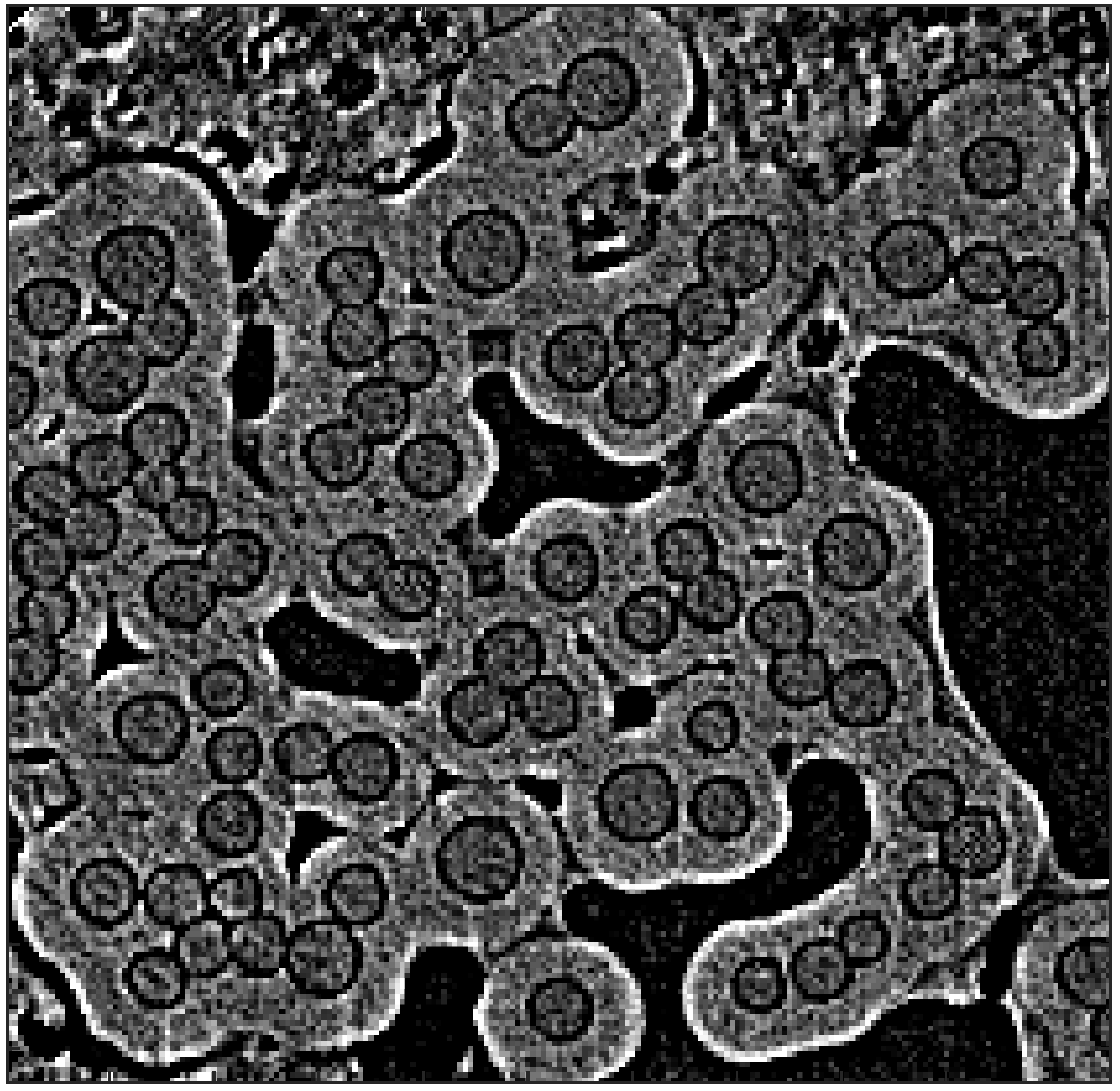}} 
\subfigure[MACE Reconstruction, $N=8$]{\includegraphics[width =0.22\textwidth]{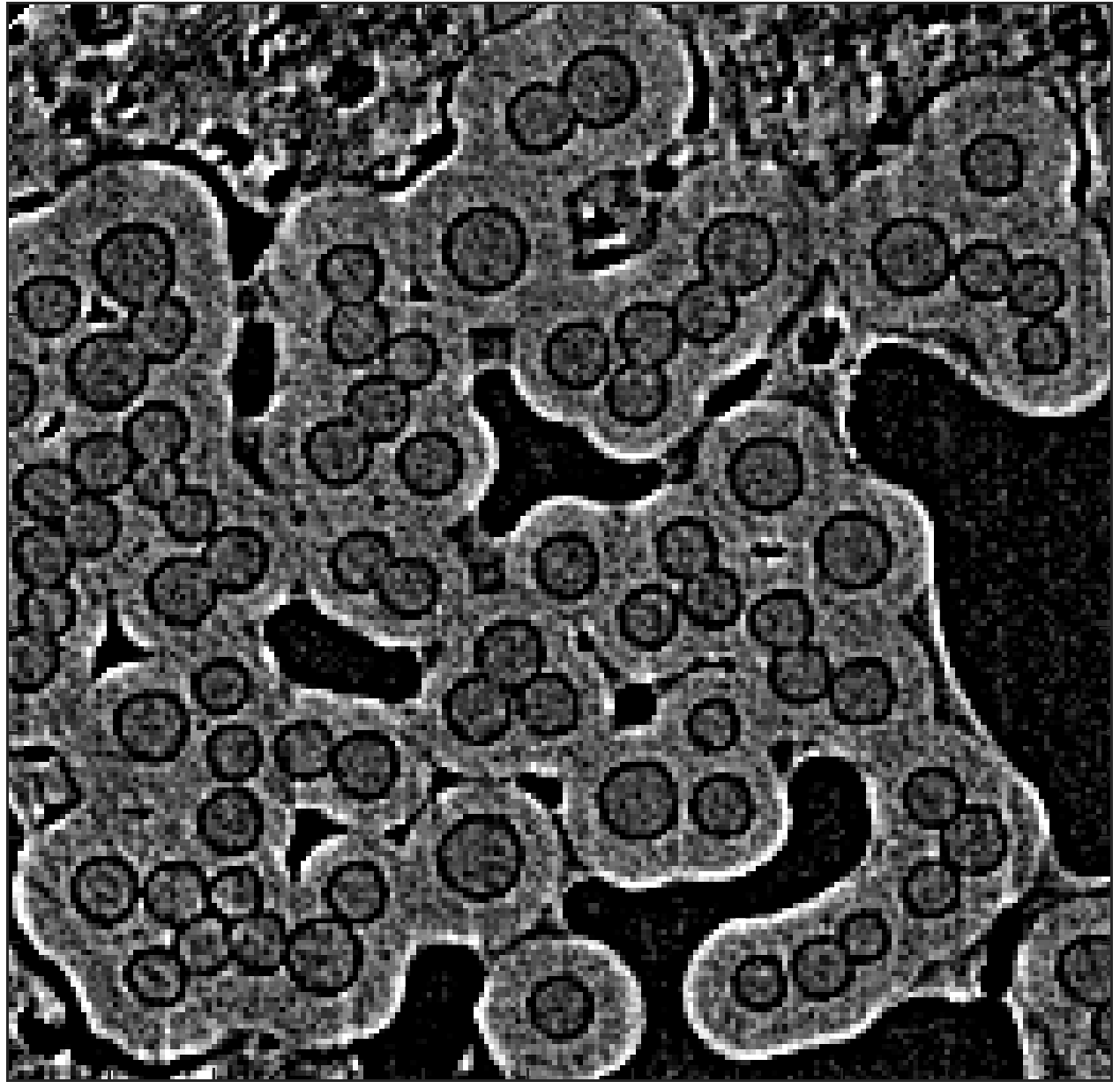}}
\caption{
Comparison of quality between (a) fully converged result and (b) the MACE reconstruction using 8 view-subsets on the NERSC supercomputer.
Notice that both have equivalent image quality.
}
\label{fig:ceramic_large_zoomed}
\end{figure}

\begin{figure}
\centering
\includegraphics[width =0.40\textwidth]{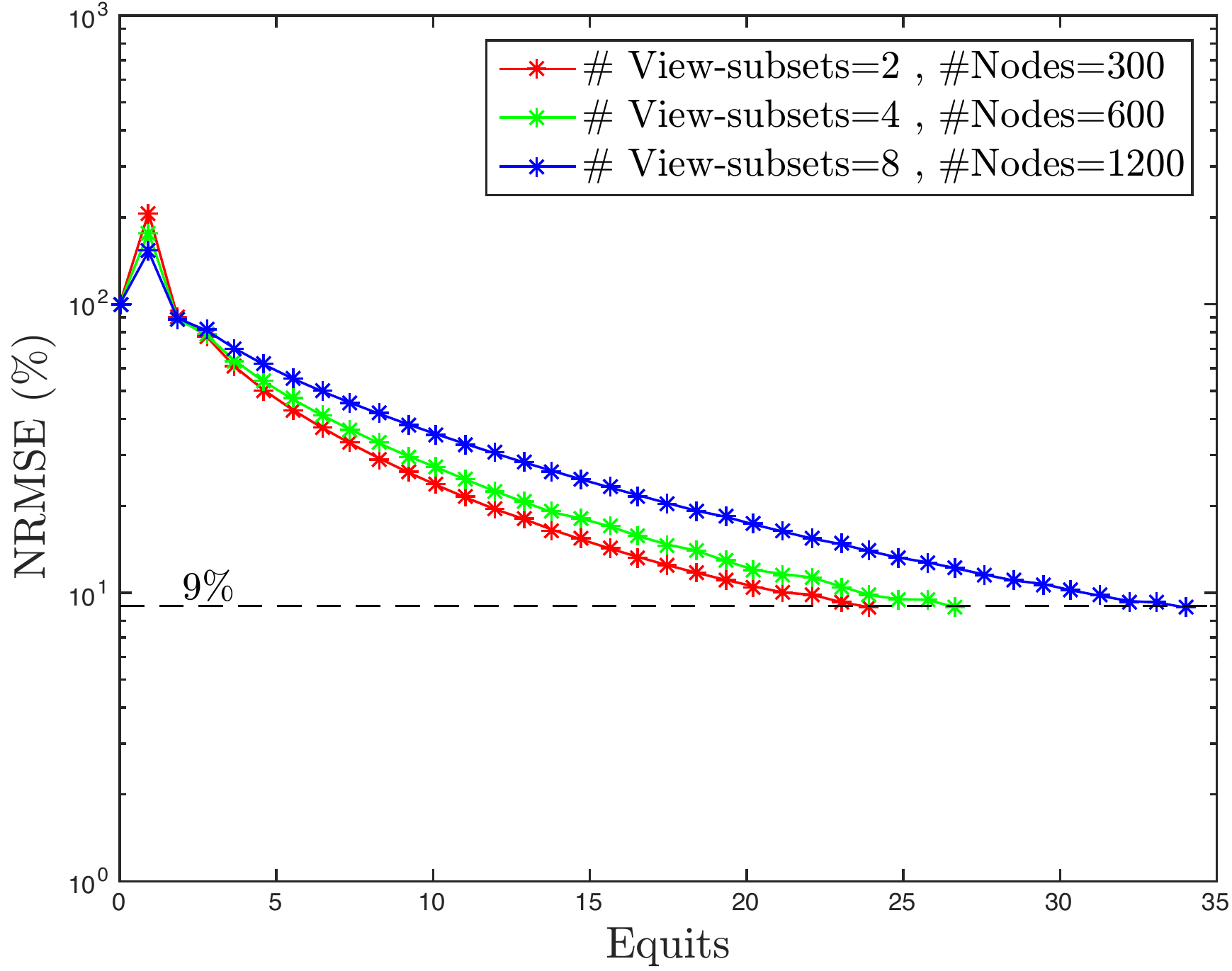} 
\caption{MACE convergence for the High-Res. ceramic reconstruction on the NERSC supercomputer as a function of view-subsets (reconstruction size $1280 \times 1280 \times 1200$).} 
\label{fig:ceramic_large_plots}
\end{figure}

\begin{table}[!ht]
\centering
\small
\begin{threeparttable}
\caption{Computational performance for 3-D MACE reconstruction on the NERSC supercomputer as a function of \#view-subsets}
\begin{tabular}{|c|c|c|c|c|}
\hline
\#View-subsets & 1 & 2 & 4 & 8\\
\hline
\#Nodes & 150 & 300 & 600 & 1200\\  
\hline
\#Cores & 10,200 & 20,400 & 40,800 & 81,600 \\
\hline
Memory usage \tnote{2} (GB) & - & 25.12 & 12.56 & 6.28 \\
\hline
\#Equits & - & 23.91 & 26.67 & 34.03\\
\hline
\#Time (s) & - & 275 & 154 & 121 \\
\hline
\shortstack{MACE Algorithmic \\ Speedup} & - & 2 & 3.58 & 5.62\\
\hline
\shortstack{Machine-time \\ Speedup} & - & 2 & 3.57 & 4.55\\
\hline 
\end{tabular}
\begin{tablenotes}
\item[2] \footnotesize{when system-matrix is represented in conventional sparse matrix format, floating-point precision} 
\end{tablenotes}
\end{threeparttable}

\label{tab:ceramic_large}
\end{table}

\subsection{MACE Reconstruction for large 3D datasets}

In this section, we study the convergence and parallel efficiency of the 3D data sets of Table~\ref{tab:Datasets}(b) using the MACE algorithm implemented on the NERSC supercomputer.

All cases use the SV-ICD algorithm for computation of the reconstructions at individual nodes with a  Q-GGMRF as prior model and $\rho=0.8$ as value of Mann parameter.
As noted previously, for this case the system matrix is so large that it is not possible to compute the reconstruction with $N=1$ view subset.
So in order to produce our reference reconstruction, we ran 40 equits of MACE with $N=2$ view subsets, which appeared to achieve full convergence.

Figure~\ref{fig:ceramic_large_zoomed} compares zoomed in regions of the fully converged result with MACE reconstructions using $N=8$ view subsets.
Notice both have equivalent image quality.

Figure~\ref{fig:ceramic_large_plots} shows the convergence of MACE as a function of number of equits. 
Notice that, as in the 2D case using the Q-GGMRF prior, the number of equits tends to increase as the number of view subsets, $N$, increases.

Table \ref{tab:ceramic_large} summarizes the results of the experiment as a function of the number of view subsets, $N$.
Notice that the memory requirements for reconstruction drop rapidly with the number of view subsets.
This makes parallel reconstruction practical for large tomographic data sets.
In fact, the results are not listed for the case of $N=1$, since the system matrix is too large to store on the nodes of the NERSC supercomputer.
Also, notice that parallel efficiency is good up until $N=4$, but it drops off with $N=8$.

\begin{appendices}

\section{}
We show that the MACE equations of (\ref{eq:CEsys_compact}) can be formulated as a fixed-point problem represented by (\ref{eq:w*_fixed_pt}). For a more detailed explanation see Corollary 3 of \cite{Buzzard2018CE}.
\begin{proof}
A simple calculation shows that for any $\v \in \RR^{nN}$, operator $\G$ defined in (\ref{eq:G_operators}) follows
$$
\G\G \v = \G \v, \text{ and so } (2\G-\I)(2\G-\I)\v = \v.
$$
Thus $2\G-\I$ is self-inverse. 
We define $\w^*$ as 
$$
\w^* = (2\G-\I)\v^*,
$$
in which case $\v^*=(2\G-\I)\w^*$ due to the above self-inverse property.
Additionally, (\ref{eq:CEsys_compact}) gives
$$
(2\F-\I)\v^* = (2\G-\I)\v^*.
$$
Note that the RHS of the above is merely $\w^*$.
So, plugging $\v^*$ in terms of $\w^*$ on the LHS, we get
\begin{align*}
(2\F-\I)(2\G-\I)\w^* = \w^*.
\end{align*}
Hence $\w^*$ fixed-point of a specific map $\T:\RR^{nN} \rightarrow \RR^{nN}$, where  $\T=(2\F-\I)(2\G-\I)$.
Finding $\w^*$ gives us $\v^*$, since $\v^*=(2\G-\I)\w^*$.
\end{proof}

\section{}

\begin{lemma}
\label{lemma:pu_icd}
Let $F_i : \RR^n \rightarrow \RR^n$, denote the proximal map of a strictly convex and continuously differentiable function $f_i:\RR^n \rightarrow \RR$.
Let $\es \in \RR^n$ be defined as $\es=[0, \cdots, 1, \cdots, 0]^t$ where entry 1 is at $s$-th index.
Let $\tilde{F}_i(v;x)$ denote the partial-update for $F_i(v)$ from an initial state $x$ as shown in Algorithm \ref{alg:pu_icd}.
Then $F_i(v)=x$ if and only if $\tilde{F}_i(v;x) =x$.
\end{lemma}

\begin{proof}
We first assume $\tilde{F}_i(v;x)=x$. 
Since $f_i$ is strictly convex and continuously differentiable,  line \ref{op:alpha_update} of Algorithm \ref{alg:pu_icd} can be re-written as
\begin{align}
\alpha_s = \left\{ \alpha \ \middle| \ \es^t \left [ \nabla f_i(z+\alpha \es) + \frac{1}{\sigma^2} (z+\alpha \es-v) \right]=0  \right\}.
\label{eq:alpha}
\end{align}
Since $\tilde{F}_i(v;x)=x$, from line \ref{op:z_update} of Algorithm \ref{alg:pu_icd} and the fact that $\es, s=1,\cdots,n$ are independent, it follows that $\alpha_s=0, s=1,\cdots, n$.
Applying $\alpha_s=0$ repeatedly to lines \ref{op:alpha_update}-\ref{op:z_update} of Algorithm \ref{alg:pu_icd} and using (\ref{eq:alpha}), we get
$$
\frac{\partial f_i(x)}{\partial x_s} + \frac{1}{\sigma^2} (x_s-v_s)=0, \ \ s=1,\cdots,n.
$$
Stacking the above result vertically, we get 
$$
\nabla f_i(x) + \frac{1}{\sigma^2} (x-v)=0.
$$
Since $f_i$ is strictly convex and continuously differentiable the above gives
$$
x = \argmin_z \left\{f_i(z) + \frac{\|z-v\|^2}{2 \sigma^2} \right\}
$$
and so, $x = F_i(v)$.
Therefore, $\tilde{F}_i(v;x)=x$ gives $F_i(v)=x$.

The converse can be proved by reversing the above steps.
\end{proof}

\begin{proof}\textbf{Proof of Theorem \ref{thm:pu_fp} }

Assume Partial-update MACE algorithm has a fixed-point $(\w^*, \X^*)$.
Then from (\ref{eq:PUCEupdate}) we get, 
\begin{align}
\X^* &= \tilde{\F}(\v^* ; \ \X^* , \sigma )
\label{eq:X_infty} \text{ and } \\
\w^* &= 2\X^{*} - \v^*,  \label{eq:w_infty} 
\end{align}
where $\v^* = (2\G-\I)\w^*$. 
So, (\ref{eq:w_infty}) can be re-written as
$$
\w^* = 2\X^* - (2\G-\I)\w^*, \text{ which gives }
$$
$$
\X^* = \G \w^*.
$$
So, $X_i^* = \bar{\w}^*$ for $i=1,\cdots,N$, and consequently, (\ref{eq:X_infty}) can be expressed as
\begin{align*}
\bar{\w}^* = \tilde{F}_i(v_i^* \ ;\bar{\w}^*, \sigma ), \ i=1,...N. 
\end{align*}
Applying Lemma \ref{lemma:pu_icd} to the above we get
$$
\bar{\w}^* = F_i(v_i^*; \ \sigma), i=1,...,N.
$$
By stacking the above result vertically, we get
$$
\G\w^* = \F(\v^*).
$$
Based on definition of $\v^*$, the above gives
$$
\G\w^* = \F (2\G-\I)\w^*.
$$
Multiplying both LHS and RHS by 2 and further subtracting $\w^*$ from both sides, we get
$$
(2\F-\I)(2\G-\I)\w^* = \w^*.
$$
Therefore, any fixed-point of the Partial-update MACE algorithm, $(\w^*,\X^*)$, is a solution to the exact MACE approach specified by (\ref{eq:w*_fixed_pt}).

\end{proof}

\begin{proof}
\textbf{Proof of Theorem \ref{thm:PU_convergence} 
(convergence of the Partial-update MACE algorithm)
} \\
We can express the function $f_i$ defined in Theorem \ref{thm:PU_convergence} more compactly.
For this purpose, let subset $J_i$ be represented as $J_i = \{k_1, k_2, \cdots k_M\}$.
Then, $f_i$ can be compactly written as 
\begin{align}
f_i(x) =  \frac{1}{2} \| \tilde{y}_i-\tilde{A}_i x\|_{\tilde{\Lambda}_i}^2 + \frac{\beta}{N} x^TBx .
\label{eq:fi_compact}
\end{align}
where $\tilde{y}_i, \tilde{A}_i$ and $\tilde{\Lambda}_i$ are defined as
$$
\tilde{y}_i =
\begin{bmatrix}
y_{k_1} \\
\vdots \\
y_{k_M}
\end{bmatrix}
\text{ , }
\tilde{A}_i =
\begin{bmatrix}
A_{k_1} \\
\vdots \\
A_{k_M}
\end{bmatrix}
\text{ and }
\tilde{\Lambda}_i =
\begin{bmatrix}
\Lambda_{k_1} & &  \\
& \ddots &  \\
& & \Lambda_{k_M}
\end{bmatrix}.
$$
From (\ref{eq:fi_compact}), we can express $F_i$, the proximal map of $f_i$, as
\begin{align}
F_i(v) &= \argmin_{z \in \RR^n} \left\{f_i(z) + \frac{\|z-v\|^2}{2 \sigma^2} \right\} \nonumber \\
&= \argmin_{z \in \RR^n} \left\{ \frac{1}{2} z^t \left( H_i + \frac{I}{\sigma^2} \right) z - z^t \left( b_i + \frac{v}{\sigma^2} \right) \right\}
\label{eq:prox_exact},
\end{align}
where $H_i \in \RR^{n \times n}$ and $b_i \in \RR^n$ are defined as  
\begin{align*}
H_i = \tilde{A}_i^t \tilde{\Lambda}_i \tilde{A}_i + (\beta/N) B  \ \text{ and } \ 
b_i = \tilde{A}_i^t \tilde{y}_i .
\end{align*}
We can obtain $\tilde{F_i}(v;x)$, the partial-update for $F_i(v)$ defined in (\ref{eq:prox_exact}), by using the convergence analysis of \cite{BoumanBook, BoumanSauerCT93} for ICD optimization.
This gives
\begin{equation}
\tilde{F}_i(v;x)= -(L_i+D_i+\sigma^{-2}I)^{-1}(L_i^tx-b_i-\sigma^{-2}v),
\label{eq:ICDUpdate1}
\end{equation}
where matrices $L_i, D_i \in \RR^{n \times n}$, are defined as
\begin{align*}
L_i &= \text{Lower triangular sub-matrix of }H_i \text{ (excluding diag.)}\\
D_i &= \text{Diagonal sub-matrix of }H_i. 
\end{align*}
Further we define $\tilde{L}_i \in \RR^{n \times n}$ by $\tilde{L}_i = L_i + D_i.$
We can re-write equation (\ref{eq:ICDUpdate1}) as 
\begin{align}
\tilde{F}_i(v;x) &= -(\tilde{L}_i+\sigma^{-2}I)^{-1}(L_i^tx-b_i-\sigma^{-2}v)  \nonumber \\
&= -\sigma^2 (I + \sigma^{2} \tilde{L}_i)^{-1}(L_i^tx-b_i-\sigma^{-2}v)
\label{eq:ICDUpdate2} 
\end{align}
Let $\varepsilon = \sigma^2$. 
For sufficiently small $\varepsilon^2$, we can approximate $(I + \varepsilon \tilde{L}_i)^{-1}$ in \eqref{eq:ICDUpdate2} as
$$
(I + \varepsilon \tilde{L}_i)^{-1} = I - \varepsilon \tilde{L}_i + \mathcal O(\varepsilon^2).
$$ 
Plugging the above approximation into equation (\ref{eq:ICDUpdate2}), simplifying, and dropping the $\mathcal O(\varepsilon^2)$ term, we get
\begin{align}
\tilde{F}_i(v;x) = (I - \varepsilon \tilde{L}_i)v-\varepsilon L_i^t x + \varepsilon b_i
\label{eq:ICDUpdate3}
\end{align}
and hence
\begin{align}
2\tilde{F}_i(v;x)-v = (I - 2\varepsilon \tilde{L}_i)v-2\varepsilon L_i^t x +2 \varepsilon b_i.
\label{eq:ReflectedICDUpdate}
\end{align}

Define block matrices $\Ls \in \RR^{Nn \times Nn}$ and $\Lt \in \RR^{Nn \times Nn}$ with $L_i$ and ${\tilde L}_i$ along the diagonal, respectively.

Using (\ref{eq:ICDUpdate3}), (\ref{eq:ReflectedICDUpdate}), $\Ls$, and $\Lt$ in (\ref{eq:PUCEupdate}), we can express the Partial-update MACE update up to terms involving $\mathcal O(\varepsilon^2)$ as
\begin{alignat}{1}
&\begin{bmatrix}
\w^{(k+1)} \\
X^{(k+1)} \\
\end{bmatrix}
=
\begin{bmatrix}
\rho & 0 \\
0 & 1 \\
\end{bmatrix}
\begin{bmatrix}
\I-2\varepsilon \Lt & -2\varepsilon \Ls^t \\
\I-\varepsilon \Lt & -\varepsilon \Ls^t  \\
\end{bmatrix}
\begin{bmatrix}
\vs^{(k)} \\
X^{(k)} \\
\end{bmatrix} \nonumber \\
& \mbox{\hspace{1in}}
\begin{bmatrix}
1-\rho & 0 \\[2mm]
0 & 0 \\
\end{bmatrix}
\begin{bmatrix}
\w^{(k)} \\
X^{(k)} \\
\end{bmatrix}
+
\cs
\nonumber \\
&=
\begin{bmatrix}
\rho & 0 \\
0 & 1 \\
\end{bmatrix}
\begin{bmatrix}
\I-2\varepsilon \Lt & -2\varepsilon \Ls^t \\
\I-\varepsilon \Lt & -\varepsilon \Ls^t  \\
\end{bmatrix}
\begin{bmatrix}
2\G-\I & \0 \\
\0 & \I \\ 
\end{bmatrix}
\begin{bmatrix}
\w^{(k)} \\
X^{(k)} \\
\end{bmatrix} \nonumber \\
&\mbox{\hspace{1in}}+
\begin{bmatrix}
1-\rho & 0 \\
0 & 0 \\
\end{bmatrix}
\begin{bmatrix}
\w^{(k)} \\
X^{(k)} \\
\end{bmatrix} \nonumber
+
\cs
\nonumber \\
&=
\begin{bmatrix}
\rho(\I-2\varepsilon \Lt)(2\G-\I)+(1-\rho)\I & -2\rho\varepsilon\Ls^t \\
(\I-\varepsilon \Lt)(2\G-\I) & -\varepsilon \Ls^t \\
\end{bmatrix} \nonumber \\
&\mbox{\hspace{1in}} \times
\begin{bmatrix}
\w^{(k)} \\
X^{(k)} \\
\end{bmatrix} 
+\cs 
\label{eq:PUCESys},
\end{alignat}
where $\cs \in \RR^{2nN}$ is a constant term based on variables $\rho$ and $b_i$.
Define $\M_\varepsilon \in \RR^{2nN \times 2nN}$ and  $\z \in \RR^{2nN}$ as follows
\begin{align}
\M_\varepsilon
&=
\begin{bmatrix}
\rho(\I-2\varepsilon \Lt)(2\G-\I)+(1-\rho)\I & -2\rho\varepsilon\Ls^t \\
(\I-\varepsilon \Lt)(2\G-\I) & -\varepsilon \Ls^t \\
\end{bmatrix} 
\label{eq:Tdef} \\
\z
&=
\begin{bmatrix}
\w \\
X \\
\end{bmatrix}
\label{eq:zdef}
\end{align}
Then we can re-write (\ref{eq:PUCESys}) as follows
\begin{align}
\z^{(k+1)}=\M_\varepsilon \z^{(k)}+\cs
\label{eq:zsys}
\end{align}
For $\z^{(k)}$ to converge in the limit $k \rightarrow \infty$, the absolute value of any eigenvalue of $\M_\varepsilon$ must be in the range $(0,1)$.
We first determine the eigenvalues of $\M_0$, where $\M_0 = \text{limit}_{\varepsilon \rightarrow 0} \M_{\varepsilon}$, and then apply a 1st order approximation in $\varepsilon$ to obtain the eigenvalues of $\M_\varepsilon$.
We can express $\M_0$ as
$$
\M_0 = 
\begin{bmatrix}
2\rho\G+(1-2\rho)\I & 0 \\
2\G-\I & 0 \\
\end{bmatrix}
$$

Let $\lambda_0 \in \RR$ and $\z_0 = [\w_0^t \ X_0^t]^t \in \RR^{2nN}$ represent eigenvalue and eigenvector of $\M_0$.
Then $\M_0 \z_0 = \lambda_0 \z_0 $, and so
\begin{align}
\G\w_0 &= \frac{1}{2\rho}(\lambda_0+2\rho-1)\w_0 
\label{eq:T0_eigenvalue1} \\
(2\G-\I)\w_0 &=\lambda_0 X_0 
\label{eq:T0_eigenvalue2}
\end{align}

Since $\G$ is an orthogonal projection onto a subspace, all of its eigenvalues are 0 or 1. 
This with (\ref{eq:T0_eigenvalue1}) implies that $\lambda_0 + 2 \rho - 1$ is 0 or $2\rho$.  
In the first case, $\lambda_0 = 1 - 2 \rho$, which lies in the open interval $(-1,1)$ for $\rho$ in $(0,1)$.  
In the second case, $\lambda_0 = 1$.
Applying this in (\ref{eq:T0_eigenvalue1}) and (\ref{eq:T0_eigenvalue2}), we get
$X_0 = \w_0=\G\w_0$, so that each eigenvector for $\lambda_0 = 1$ has $X_0 = \w_0$ with all subvectors identical.  

Let $\lambda_{\varepsilon} \in \RR$ and $\z_{\varepsilon} = [\w_{\varepsilon}^t \ X_{\varepsilon}^t]^t \in \RR^{nN}$ represent eigenvalue and eigenvector of $\M_{\varepsilon}$ respectively.
Let $a$ be the derivative of $\lambda_\varepsilon$ with respect to $\varepsilon$, and let $\u_1=\nabla_{\varepsilon} \w_\varepsilon$ and $\u_2=\nabla_{\varepsilon} X_\varepsilon$.
Applying a 1st order approximation in $\varepsilon$, $\lambda_{\varepsilon}$ and $\z_{\varepsilon}$ are given by
\begin{align*}
\lambda_{\varepsilon} &= \lambda_0 + a(\varepsilon-0) = 1 + a\varepsilon \\
\z_{\varepsilon} 
&= 
\begin{bmatrix}
\w_0 + (\varepsilon-0) \u_1 \\
X_0 + (\varepsilon-0) \u_2 \\
\end{bmatrix}
=
\begin{bmatrix}
\w_0 + \varepsilon \u_1 \\
\w_0 + \varepsilon \u_2 \\
\end{bmatrix}.
\end{align*}
If we can prove that $a$ is negative when $\varepsilon$ is infinitely small positive,
then consequently $|\lambda_\varepsilon| < 1$, and so, the system of equations specified by equation (\ref{eq:PUCESys}) converges (note that the case of $\lambda_0 = 1- 2 \rho$ gives $|\lambda_\varepsilon| < 1$ by continuity for small $\varepsilon$).
Since $\M_\varepsilon \z_\varepsilon = \lambda_\varepsilon \z_\varepsilon$, the first component of equation (\ref{eq:Tdef}) gives
\begin{align*}
& \left[ \rho(\I - 2\varepsilon \Lt)(2\G-\I)+(1-\rho)\I \right](\w_0+\varepsilon \u_1) \\
&\mbox{\hspace{0.5in}}- 2\rho\varepsilon\Ls^t(\w_0+\varepsilon \u_2) 
 = (1+a \varepsilon)(\w_0+\varepsilon \u_1).
\end{align*}
Neglecting terms $\mathcal O(\varepsilon^2)$, expanding, and using $(2\G-\I)\w_0=\w_0$, the above simplifies for $\varepsilon > 0$ to 
$$
2\rho(\G-\I)\u_1 = \left [   2\rho(\Lt+\Ls^t) + a\I \right ]\w_0.
$$
Applying $\G$ to both sides, using $\G(\G-\I)=\G^2 - \G = 0$ and $\G\w_0=\w_0$, we get
$$
\0 = 2\rho \G(\Lt+\Ls^t)\w_0 + a\w_0 
$$
and so,
\begin{align}
\G(\Lt+\Ls^t)\w_0 = -\frac{a}{2\rho}\w_0. 
\label{eq:a_posdefinite1}
\end{align}
Since $\tilde{L}_i+L_i^t, i=1,...,N$ is positive definite for each $i$, so is  $\Hs = \Lt+\Ls^t$.
Further, $\G \in \RR^{nN \times nN}$ is an orthogonal projection matrix with $n$-dimensional range.  Hence 
$\G$ can be expressed as $\G=PP^t$, where $P \in \RR^{nN \times n}$ is orthogonal basis of the range of $\G$ (i.e $P^tP=I$). 
Since $\w_0=\G\w_0$, equation (\ref{eq:a_posdefinite1}) can be written as
$$
PP^t\Hs PP^t\w_0 = -\frac{a}{2\rho} PP^t\w_0
$$
Multiply both LHS and RHS by $P^t$, and define $\tilde{\w}_0=P^t\w_0$.
Since $P^tP=I$, we get
$$
P^t\Hs P \tilde{\w}_0 = -\frac{a}{2\rho} \tilde{\w}_0
$$
This implies that $-a/(2\rho)$ is an eigenvalue of $P^t\Hs P$.  Since $\rho > 0$ and $P^t\Hs P$ is positive definite, we have $a<0$, and consequently, $|\lambda_\varepsilon| < 1$.

Since all eigenvalues of $\M_{\varepsilon}$ have absolute value less than 1, the system of equations specified by (\ref{eq:PUCEupdate}) converges to a point $\z^*= (\w^*, \X^*)$ in the limit $k \rightarrow \infty$. 
From Theorem \ref{thm:pu_fp}, $\w^*$ is a solution to the exact MACE approach specified by (\ref{eq:w*_fixed_pt}).

\end{proof}

\section{}

\begin{theorem}
Let $F_i, i=1,\cdots,N$, be the proximal map of a closed, convex, differentiable function, $f_i$, let $\sum_{i=1}^N f_i = f$, let $F$ be the proximal map for $f$, and let $H$ be any denoiser.
Then, the MACE framework specified by equilibirum conditions (\ref{eq:ParallelPnP_F}) \textendash \  (\ref{eq:ParallelPnP_usum}) is exactly equivalent to the standard PnP framework specified by (\ref{eq:PnP_F}) and (\ref{eq:PnP_H}).
\end{theorem}

\begin{proof}
Assume (\ref{eq:ParallelPnP_F}) \textendash \  (\ref{eq:ParallelPnP_usum}) hold.
Then, as per (\ref{eq:ParallelPnP_F}),
$$
x^*=F_i(x^*+u_i^* ; \sigma), i=1,...,N.
$$
Since $f_i$ is convex, differentiable, and $F_i$ is defined as
$$
F_i(x ; \sigma) = \argmin_{v \in \RR^n} \left\{f_i(v) + \frac{\|v-x\|^2}{2 \sigma^2} \right\}, 
$$ 
it follows from the above stated equilibrium condition that
\begin{align}
\nabla f_i(x^*) + \frac{x^*-(x^*+u_i^*)}{\sigma^2} &= 0, \text{ or, } \nonumber \\
\nabla f_i(x^*) - \frac{u_i^*}{\sigma^2} &= 0.
\label{eq:Prox_deriv}
\end{align}
Summing the above equation over $i=1,...,N$ we get
$$
\sum_{i=1}^N \nabla f_i(x^*) - \frac{N \bar{\u}^*}{\sigma^2} = 0,
$$
where $\bar{\u}=\sum_{i=1}^N u_i/N$.
Since $f = \sum_{i=1}^N f_i$, the above can be re-written as
$$
\nabla f(x^*) + \frac{x^*-(x^*+N\bar{\u}^*)}{\sigma^2} = 0.
$$
Since $f$ is convex, the above equation implies that 
$$
x^* = \argmin_{v \in \RR^n} \left\{f(v) + \frac{\|v-(x^*+N\bar{\u}^*)\|^2}{2 \sigma^2} \right\}, \text{ and so, }
$$
$$
x^* = F(x^*+N\bar{\u}^*; \sigma).
$$
From (\ref{eq:ParallelPnP_H}) and (\ref{eq:ParallelPnP_usum}), we additionally get $x^*=H(x^*-N\bar{\u}^*)$.
Therefore, we get (\ref{eq:PnP_F}) and (\ref{eq:PnP_H}), where $\alpha^*=-N\bar{\u}^*$.

For the converse, assume (\ref{eq:PnP_F}) and (\ref{eq:PnP_H}) hold.
Then, as per (\ref{eq:PnP_F}), $F(x^*-\alpha^*; \sigma) = x^*$.
So, we get
$$
\nabla f(x^*) + \frac{x^*-(x^*-\alpha^*)}{\sigma^2}=0.
$$
Since $f=\sum_{i=1}^N f_i$, we can re-write the above as
$$
\alpha^* = -\sum_{i=1}^N \sigma^2 \nabla f_i(x^*).
$$
We define $u_i^*$ as $u_i^* = \sigma^2 \nabla f_i(x^*)$.
So, from the above equation, we get
$
\alpha^* + \sum_{i=1}^N u_i^*=0 ,
$
which gives (\ref{eq:ParallelPnP_usum}).
Further from the defintion of $u_i^*$ we get
$$
\nabla f_i(x^*) + \frac{x^*-(x^*+u_i^*)}{\sigma^2}=0, \text{ and so, }
$$
$$
F_i(x^*+u_i^*; \sigma)=x^*,
$$
which gives (\ref{eq:ParallelPnP_F}).
Also, as per (\ref{eq:PnP_H}), $H(x^* + \alpha^*)=x^*$, which gives (\ref{eq:ParallelPnP_H}).
Therefore, we obtain (\ref{eq:ParallelPnP_F}) \textendash \ (\ref{eq:ParallelPnP_usum}), where $u_i^*=\sigma^2 \nabla f_i(x^*), i=1,\cdots,N$.
\end{proof}

{\em Remark: } As in \cite{Buzzard2018CE}, the theorem statement and proof can be modified to allow for nondifferentiable, but still convex functions, $f_i$.
\medskip

\begin{proof} \textbf{Proof of Theorem \ref{thm:FixedPtSolutionToPnP} } \\ Assume (\ref{eq:ParallelPnP_F}) \textendash \  (\ref{eq:ParallelPnP_usum}) hold.
We define $\t^*$ as $\t^*=N\u^*$. 
So, (\ref{eq:ParallelPnP_usum}) gives
$
\alpha^* +(\sum_{i=1}^N t_i^*)/N=0, \text{or, } \alpha^* = -\bar{\t}^*. 
$
Consequently, we can express (\ref{eq:ParallelPnP_H}) as
\begin{align}
H(x^*-\bar{\t}^*)=x^*.
\label{eq:H_xt}
\end{align}
Further, (\ref{eq:ParallelPnP_F}) specifies 
$
F_i(x^*+u_i^*; \sigma)=x^*.
$ 
We showed earlier in (\ref{eq:Prox_deriv}) that this gives $\nabla f_i(x) - u_i/\sigma^2 = 0$.
So, we get
\begin{align}
\nabla f_i(x^*) - \frac{N u_i^*}{N \sigma^2} &= 0, \ \text{ or, } \nonumber \\
\nabla f_i(x^*) + \frac{x^* - (x^*+t_i^*)}{ (\sqrt{N}\sigma)^2} &= 0, \ \text{ or, } \nonumber \\
F_i(x^*+t_i^*; \sqrt{N} \sigma) &= x^*, \ i=1,...,N. \label{eq:F_xt}
\end{align}
Define $\w^*$ as $\w^*=\hat{x}^*-\t^*$, where $\hat{x}^*$ is $N$ vertical copies of $x^*$. 
We write (\ref{eq:H_xt}) as $\G_H\w^*=\hat{x}^*$.
So, based on definition of $\w^*$, we have $\t^*=\hat{x}^*-\w^*= \G_H\w^*-\w^* = (\G_H-\I)\w^*$.
We can write (\ref{eq:F_xt}) as $\F(\hat{x}^*+\t^*)=\hat{x}^*$ 
according to (\ref{eq:FG_PnP})
, and so, by plugging in $\hat{x}^*$ and $\t^*$ in terms of $\w^*$ we get
\begin{align*} 
\F(\G_H \w^* + (\G_H-\I) \w^*) &= \G_H \w^*, \text{ or}, \\
\F(2\G_H-\I)\w^* &= \G_H \w^*.
\end{align*}
Multiplying by 2 and adding $\w^*$ on both sides we get
$$
(2\F-\I)(2\G_H-\I)\w^* = \w^*.
$$
Therefore, $\w^*$ is a fixed-point of $\T_H=(2\F-\I)(2\G_H-\I)$, and, $\G_H(\w^*)=\hat{x}^*$, where $\w^*$ is given by $\w^*=\hat{x}^*-N\u^*$.

For the converse, assume $(2\F-\I)(2\G_H-\I)\w^*=\w^*$ and $\G_H\w^*=\hat{x}^*$ hold.
The former gives $\F(2\G_H-\I)\w^*=\G_H\w^*$. 
Applying the latter, we get $\F(2\hat{x}^*-\w^*)=\hat{x}^*$.
Define $\t^*$ as $\t^*=\hat{x}^*-\w^*$.
So, we have $\F(\hat{x}^*+\t^*)=\hat{x}^*$, or,
$$ F_i(x^*+t_i^*; \sqrt{N}\sigma)=x^*, \ i=1,\cdots,N.$$
A calculation with the definition of $F_i$ shows that the above gives
$$ F_i(x^*+t_i^*/N; \sigma) = x^*,  \ i=1,\cdots,N. $$
Define $\u^* = \t^*/N$.
So, from the above, $F_i(x^*+u_i^*; \sigma)=x^*$, which gives (\ref{eq:ParallelPnP_F}).
Since $\G_H\w^*=\hat{x}^*$, we have $x^*=H\bar{\w}^*$. 
Combining this with $\w^*=\hat{x}^*-\t^*$, we get $x^*=H(x^*-\bar{\t}^*)$.
Define $\alpha^* = -\bar{\t}^*$. 
So we have,  $x^*=H(x^*+\alpha^*)$, which gives (\ref{eq:ParallelPnP_H}).
Also from definition of $\alpha^*$ and $\u^*$, we get $\alpha^* + \sum_{i=1}^N u_i^* = 0$, which gives (\ref{eq:ParallelPnP_usum}). 
Therefore, we obtain (\ref{eq:ParallelPnP_F}) to (\ref{eq:ParallelPnP_usum}), where $\u^*$ is given by $N\u^*=\hat{x}^*-\w^*$.

\end{proof} 


\begin{proof}\textbf {Proof of Lemma \ref{lem:HG}} \\
First we show that $2\G_H-\I$ is non-expansive when $H$ is firmly non-expansive. 
This proof also applies to the case where $H$ is a proximal map, since proximal maps are firmly non-expansive \cite{ParikhProximal}. 
Consider any $\x,\y \in \RR^{nN}$.
Then 
\begin{align*}
&\|(2\G_H-\I)\x-(2\G_H-\I)\y\|^2 \\
& = 4\|\G_H\x-\G_H\y\|^2+\|\x-\y\|^2 
- 4\<\G_H\x-\G_H\y,\x-\y\>
\label{eq:GH_norm}
\end{align*}
By writing $\G_H$ in terms of $H$, we simplify the last term as
\begin{flalign*}
\<\G_H\x&-\G_H\y,\x-\y\> \\
&= \sum_{i=1}^{N} \<H\bar{\x}-H\bar{\y}, \  x_i-y_i\> \\
&= \sum_{i=1}^{N} \<H\bar{\x}-H\bar{\y}, \ \bar{\x}-\bar{\y}+(x_i-\bar{\x})-(y_i-\bar{\y})\> \\
&= N \<H\bar{\x}-H\bar{\y}, \ \bar{\x}-\bar{\y}\>.
\end{flalign*}

Since $H$ is \emph{firmly non-expansive},
\cite[Prop.~4.2]{Bauschke2011} implies that $\<H\bar{\x}-H\bar{\y}, \ \bar{\x}-\bar{\y}\> \ge \|H\bar{\x}-H\bar{\y}\|^2$.
This gives
$$
\<\G_H\x-\G_H\y,\x-\y\> \ge N\|H\bar{\x}-H\bar{\y}\|^2 = \|\G_H\x-\G_H\y\|^2.
$$
Plugging the above into the first equation of this proof, we get
$$
\|(2\G_H-\I)\x-(2\G_H-\I)\y\|^2 \le \|\x-\y\|^2.
$$
Therefore, $(2\G_H-\I)$ is a non-expansive map. 
Also, since $F_i, i=1,...,N$ is the proximal map of a convex function, $F_i$ is a resolvent operator, so $2F_i-I$ is a reflected resolvent operator, hence non-expansive. This means $2\F-\I$ is non-expansive, so $(2\F-\I)(2\G_H-\I)$ is non-expansive, since it is the composition of two non-expansive maps.
\end{proof}

\end{appendices}

\section*{Acknowledgements}
We thank the Awareness and Localization of Explosives-Related Threats (ALERT) program, Northeastern University for providing us with baggage scan dataset used in this paper.
We also thank Prof. Michael W. Czabaj, University of Utah for providing us both the Low-Res. and High-Res. ceramic composite datasets used in this paper.

This work was supported partly by the US Dept. of Homeland Security, S\&T Directorate, under Grant Award 2013-ST-061-ED0001
and by the NSF under grant award  CCF-1763896.
The views and conclusions contained in this document are those of the authors and should not be interpreted as necessarily representing the official policies, either expressed or implied, of the U.S. Department of Homeland Security.

\bibliographystyle{IEEEtran}
\bibliography{References}

\end{document}